\documentclass[apj,twocolumn, twocolappendix]{openjournal}
\usepackage{orcidlink}  

\usepackage{multirow}
\usepackage[parfill]{parskip}
\usepackage{graphicx}
\usepackage{amssymb}
\usepackage{amsmath}
\usepackage{epstopdf}
\usepackage{xcolor}

\usepackage{hyperref}
\hypersetup{
     colorlinks   = true,
     citecolor    = blue,
     linkcolor = blue,
     urlcolor = blue
}

\def\apj{{ ApJ}}
\def\apjl{{ApJL}}
\def\apjs{{ ApJS}}

\def\aap{{ A\&A}}

\def\mnras{{ MNRAS}}
\def\araa{{ ARA\&A}}

\def\nat {{ Nature}}

\def\mr{\mathrm}
\def\d{\mr{d}}

\def\Msun{M_{\odot}}
\def\Rsun{R_{\odot}}

\def\mproton{m_{\rm p}}

\def\kB{k_{\rm B}}

\def\rd{r_{\rm d}}

\newcommand{\lrb}[1]{\left({#1}\right)}
\newcommand{\lrsb}[1]{\left[{#1}\right]}

\newcommand{\lara}[1]{\left\langle{#1}\right\rangle}

\def\msunyr{M_\odot\mr{\,yr^{-1}}}
\def\kap{\kappa_{0.2}}
\def\Mten{M_{10}}
\def\mstar{m_*}
\def\rstar{r_*}

\begin{document}


\title{Implications of the UV/optical Plateau of AT2018cow}

\vspace{-1cm}

\author{Wenbin Lu\,\orcidlink{0000-0002-1568-7461}$^{1}$}
\author{Anthony L. Piro\,\orcidlink{0000-0002-8266-0293}$^{2}$}

\affiliation{$^1$Department of Astronomy and Theoretical Astrophysics Center, University of California, Berkeley, CA 94720-3411, USA}
\affiliation{$^2$The Observatories of the Carnegie Institution for Science, Pasadena, CA 91101, USA}

\email{wenbinlu@berkeley.edu}

\begin{abstract}

AT2018cow, the prototypical luminous fast blue optical transient (LFBOT),
shows a slowly fading plateau in the UV/optical light lasting at least five years after
the initial explosion. The plateau SED is blackbody-like with
color temperature $T_{\rm BB}\gtrsim2\times10^4\,\rm K$ and an emission
radius $R_{\rm BB}\sim 40 R_\odot$, which has been attributed to a geometrically thin accretion disk around a compact object. Viscously powered disk models, however, require an unusually massive black hole $M\gtrsim 200 M_\odot$ (for viscosity parameter $\alpha\gtrsim0.01$). Moreover, accretion onto such a massive black hole produces X-ray emission that is likely in tension with the XMM-Newton constraint, unless the X-ray is highly anisotropic or absorbed. We propose instead that the UV/optical emission arises from the photosphere of a super-Eddington wind launched from the inner disk, together with irradiation and reprocessing by the geometrically thin outer disk. Fitting the model to four-band HST photometry at $t\simeq 4\,\rm yr$, we find that all accretors from $1.4\,M_\odot$ neutron star to $\sim\!100\,M_\odot$ black hole produce similarly good fits.
This degeneracy arises because the wind photospheric color temperature depends only weakly on the accretor mass $M$. We also find that, at NIR--MIR wavelengths ($\lambda\sim1$--$10\,\mu\rm m$), wind free-free emission differs by an order of magnitude across the allowed mass range, providing a potential discriminant accessible to JWST. Another potential signature to differentiate the accretor mass and disk composition is the recombination line emission (HeII1640 or H$\alpha$) from the disk wind. The inferred disk parameters from our model are consistent with a micro-tidal disruption event --- either a tidal disruption of a main-sequence companion star or a merger with a companion --- in which an initially geometrically thick disk transitions to a thin phase within the first year and subsequently the thin outer disk evolves on the observed $\sim\!10\,\rm yr$ viscous timescale.


\end{abstract}

\maketitle  

\section{Introduction}\label{sec:intro}

Luminous fast blue optical transients (LFBOTs) are a class of extragalactic
transients characterized by fast rise times ($\lesssim$ few days), high peak
optical luminosities ($\gtrsim10^{44}\rm\,erg\,s^{-1}$), and blue, featureless
spectra indicative of a hot ($T\gtrsim10^4\rm\,K$) continuum source
\citep[e.g.,][]{2014ApJ...794...23D, 2018ApJ...865L...3P, 2019ApJ...872...18M, 2019ApJ...879L..13M, 2019MNRAS.484.1031P, 2019ApJ...871...73H}. About a dozen members of the class are known to date \citep{2020ApJ...895...49H, 2020ApJ...895L..23C, 2021MNRAS.508.5138P, 2022ApJ...934..104Y, 2022ApJ...926..112B, 2023Natur.623..927H, 2025ApJ...995..228S, 2026ApJ...997L..10L, 2025ApJ...993L...6N, 2026arXiv260118926S}.
They are further distinguished by luminous X-ray and radio counterparts,
likely reflecting a central engine that drives fast ($\sim\!0.1c$) outflows and prolonged energy injection into a low-mass supernova (SN) ejecta \citep[e.g.,][]{2019ApJ...872...18M, 2019ApJ...871...73H} --- but the nature of the engine remains debated \citep{2019MNRAS.487.5618L, 2019MNRAS.484.4972S, 2022ApJ...932...84M, 2025ApJ...986...84T, 2025arXiv251009745K, 2025arXiv251209017W, 2026arXiv260118887G}.

The prototypical LFBOT AT2018cow, located at $D_{\rm L}\approx 60\rm\,Mpc$, is the most nearby and best-studied member of the class, with multi-wavelength follow-up spanning X-ray to radio wavelengths over several years. At late times ($t\gtrsim1\rm\,yr$), Hubble Space Telescope (HST) observations of AT2018cow reveal that, after an initial period of steep decay, the lightcurve flattens to a plateau lasting at least five years \citep{2023ApJ...955...43C, 2023MNRAS.519.3785S, 2023MNRAS.525.4042I, 2025MNRAS.544L.108I}. The spectral energy distribution (SED) during the plateau phase is blackbody-like with $T_{\rm BB}\gtrsim2\times10^4\rm\,K$, placing the UV/optical bands near the Rayleigh--Jeans (RJ) tail.
The characteristic emission radius inferred from the SED,
$R_{\rm BB}\sim40\,R_\odot$, is far too compact for ongoing
CSM interaction and is instead naturally explained by emission from the
outer regions of a geometrically thin accretion disk
\citep{2023ApJ...955...43C, 2024ApJ...963L..24M}.
Time-dependent disk evolution models have been used to fit the plateau
\citep{2025MNRAS.544L.108I, 2025arXiv251209017W}, and in this framework
the slow decay of the plateau implies a viscous timescale
$t_{\rm vis}\sim10\rm\,yr$ for the outer disk.

A critical question is: what is the mass of the accretor?
In the pure thin-disk picture the observed UV/optical luminosity at
$r_{\rm d}\sim40\,R_\odot$ requires the product of the accretion rate and the accretor mass to be $\dot{M}M\sim 1\,M_\odot^2\rm\,yr^{-1}$ (see \S\ref{sec:thin_disk}). For a relatively low accretor mass $M\lesssim 10M_\odot$, the corresponding high accretion rate means that the outer disk would be radiation-pressure dominated and geometrically thick (\S\ref{sec:thin_disk}). Demanding self-consistency forces the accretor mass to be
$M\gtrsim200\,M_\odot$ for viscous parameter $\alpha>0.01$, as has been found by \citet{2025MNRAS.544L.108I, 2025arXiv251209017W}.
Such black holes (BHs) sit above the pair-instability mass gap
\citep{2017ApJ...836..244W, 2019ApJ...887...53F}, motivating the consideration of rare formation channels such as direct collapse of very massive stars \citep{2026A&A...706A.327C} or tidal disruption events (TDEs) by intermediate-mass BHs \citep{2019MNRAS.484.1031P, 2019MNRAS.487.2505K}. Furthermore, XMM-Newton observations set an upper limit on the soft X-ray luminosity \citep{2024ApJ...963L..24M} that independently constrains $M\lesssim100\,M_\odot$ in accretion-powered models
(\S\ref{sec:mass_constraints}). The joint constraints leave only a narrow (if any) window for the thin-disk interpretation.

In this paper we propose a new model for the UV/optical
plateau. When the \textit{global} accretion rate exceeds the Eddington value (which is the case
for the inferred parameters), the inner regions of the disk are
geometrically thick and launch a powerful optically thick wind.
We show that emission from the wind photosphere, together with irradiation and
reprocessing by the geometrically thin outer disk, can reproduce the
UV/optical plateau for any accretor mass from $1.4\,M_\odot$ (neutron
star, NS) to $\sim100\,M_\odot$ (stellar-mass BH).
The UV/optical data alone cannot discriminate between these masses, because
the wind photospheric color depends only weakly on the accretor's mass $M$.
The mass degeneracy may, however, be broken at NIR-MIR wavelengths by
wind free-free emission, which is directly accessible with James Webb Space Telescope (JWST).

This paper is organized as follows.
In \S\ref{sec:thin_disk} we review the thin-disk-only model and its
mass constraint.
In \S\ref{sec:wind_model} we introduce the wind+irradiation model.
In \S\ref{sec:SED_fits} we fit the model to HST data and present the posteriors
and mass constraints.
In \S\ref{sec:micro-TDEs} and \S\ref{sec:comparison} we discuss the
micro-TDE disk formation scenario and compare our results with
previous large-BH-mass interpretations.
We summarize in \S\ref{sec:summary}.
The Appendix presents the full posterior distributions of our fits.

\section{Models}

We describe two models: a conventional one \citep{1973A&A....24..337S} that includes only accretion power in a geometrically thin disk, and a more complete one that includes the effects of a super-Eddington wind launched from the inner regions of the accretion disk.

\subsection{Accretion power from a thin disk}
\label{sec:thin_disk}


The measured spectral luminosity near $\nu \simeq 5.6\times10^{14}\rm\, Hz$
(HST F555W) at $t = 1453\rm\, d$ is $\nu L_\nu \simeq 3.7\times 10^{38}\rm\, erg\, s^{-1}$ \citep{2023ApJ...955...43C, 2025MNRAS.544L.108I}. Hereafter, we use $\nu_{14.75} = \nu/5.6\times10^{14}\rm\, Hz$ (normalized to the pivot frequency of the HST F555W band). In the RJ limit, the characteristic emission radius is
\begin{equation}\label{eq:R_BB}
\begin{split}
     R_{\rm BB} = &\lrb{L_\nu c^2\over 8\pi^2\nu^2 \kB T_{\rm BB}}^{1/2}
    \simeq 42\,\Rsun\, \nu_{14.75}^{-3/2}\times\\
    &\lrb{\nu L_\nu\over 3.7\times 10^{38}\rm\,erg\,s^{-1}}^{1/2}
    \lrb{T_{\rm BB}\over 2\times10^4\rm\,K}^{-1/2},
\end{split}
\end{equation}
where we have assumed an isotropic blackbody emitter with surface flux 
$\pi B_\nu \approx 2\pi\nu^2\kB T_{\rm BB}/c^2$.
This compact size of $R_{\rm BB}\sim 40R_\odot$ rules out an origin 
in ongoing circumstellar medium (CSM) interaction, which would occur at much larger radii where 
the system is optically thin at these late epochs 
\citep[e.g.,][]{2019ApJ...871...73H}.
Instead, it is naturally consistent with emission from the outer edge of a 
geometrically thin accretion disk of radius $r_{\rm d} \sim R_{\rm BB}$, 
as proposed by several previous studies
\citep{2023ApJ...955...43C, 2024ApJ...963L..24M, 2025MNRAS.544L.108I, 
2025arXiv251209017W}.

We model the outer disk as a geometrically thin, optically thick disk 
\citep{1973A&A....24..337S} with a constant accretion rate $\dot{M}_{\rm d}$ at 
radii $r_{\rm in} \leq r \leq \rd$, where $\rd$ is the outer disk radius 
and $r_{\rm in}$ is the inner truncation radius (which has negligible effect on the 
UV/optical emission).
The disk effective temperature profile is
\begin{equation}
    \sigma_{\rm SB} T_{\rm eff}^4(r) = {3GM\dot{M}_{\rm d}\over 8\pi r^3},
    \label{eq:Teff}
\end{equation}
where $\sigma_{\rm SB}$ is the Stefan-Boltzmann constant, $G$ is the gravitational constant, and $M$ is the mass of the accreting compact object. The spectral luminosity is given by $L_\nu = 4\pi\int_{r_{\rm in}}^{\rd} \pi B_\nu(T_{\rm eff})\, r\,{\rm d}r$ --- in this work, we ignore a weak geometric factor that depends on the observer's viewing angle with respect to the disk angular momentum axis. Since the optical bands lie on the RJ tail 
($h\nu \ll \kB T_{\rm eff}(r)$ for $r \lesssim \rd$), we use 
$B_\nu \approx 2\nu^2\kB T_{\rm eff}/c^2$ and obtain
\begin{equation}
\begin{split}
    \nu L_\nu \simeq\, &{32\pi^2\kB\over 5c^2}\,\nu^3 
    \lrb{3GM\dot{M}_{\rm d}\over 8\pi\sigma_{\rm SB}}^{1/4} r_{\rm d}^{5/4} \\
    =\, &4.0\times10^{38}\mr{\,erg\,s^{-1}}
    \nu_{14.75}^3
    \Mten^{1/4}\times\\
    &\lrb{\dot{M}_{\rm d}\over 0.1\,\msunyr}^{1/4}
    \lrb{\rd\over 40\Rsun}^{5/4}.
\end{split}
    \label{eq:nuLnu_RJ}
\end{equation}
Here $\Mten \equiv M/10\,\Msun$ and hereafter the accretor mass is normalized to $10\,\Msun$.
This expression depends strongly on $\rd$ but only weakly on $M$ and
$\dot{M}_{\rm d}$, so the observed luminosity primarily constrains $\rd$.

We now connect $\dot{M}_{\rm d}$ to the disk mass $M_{\rm d}$ via the viscous 
timescale $t_{\rm vis}$ of the outer disk:
\begin{equation}
    M_{\rm d} \simeq \dot{M}_{\rm d}\,t_{\rm vis}, \quad 
    t_{\rm vis} \simeq {1\over\alpha}\lrb{H\over\rd}^{-2}
    \sqrt{{\rd^3\over GM}},
    \label{eq:tvis}
\end{equation}
where $\alpha$ is the Shakura--Sunyaev viscosity parameter 
\citep{1973A&A....24..337S} and $H/\rd$ is the disk aspect ratio.
The slow, years-long evolution of the UV/optical plateau implies 
$t_{\rm vis} \sim 10\rm\, yr$, which gives
\begin{equation}
    {H\over\rd} \simeq 
    {\lrb{\rd^3/GM}^{1/4}\over(\alpha t_{\rm vis})^{1/2}}.
    \label{eq:H_over_r}
\end{equation}
We use equations~(\ref{eq:tvis}) and (\ref{eq:H_over_r}) to relate 
$M_{\rm d}$, $\rd$, $M$, $\alpha$, and $t_{\rm vis}$, and work in the 
two-dimensional $(M_{\rm d}, \rd)$ parameter space at fixed $M$ and 
$\alpha$.

\begin{figure}
\centering
\includegraphics[width=0.47\textwidth]{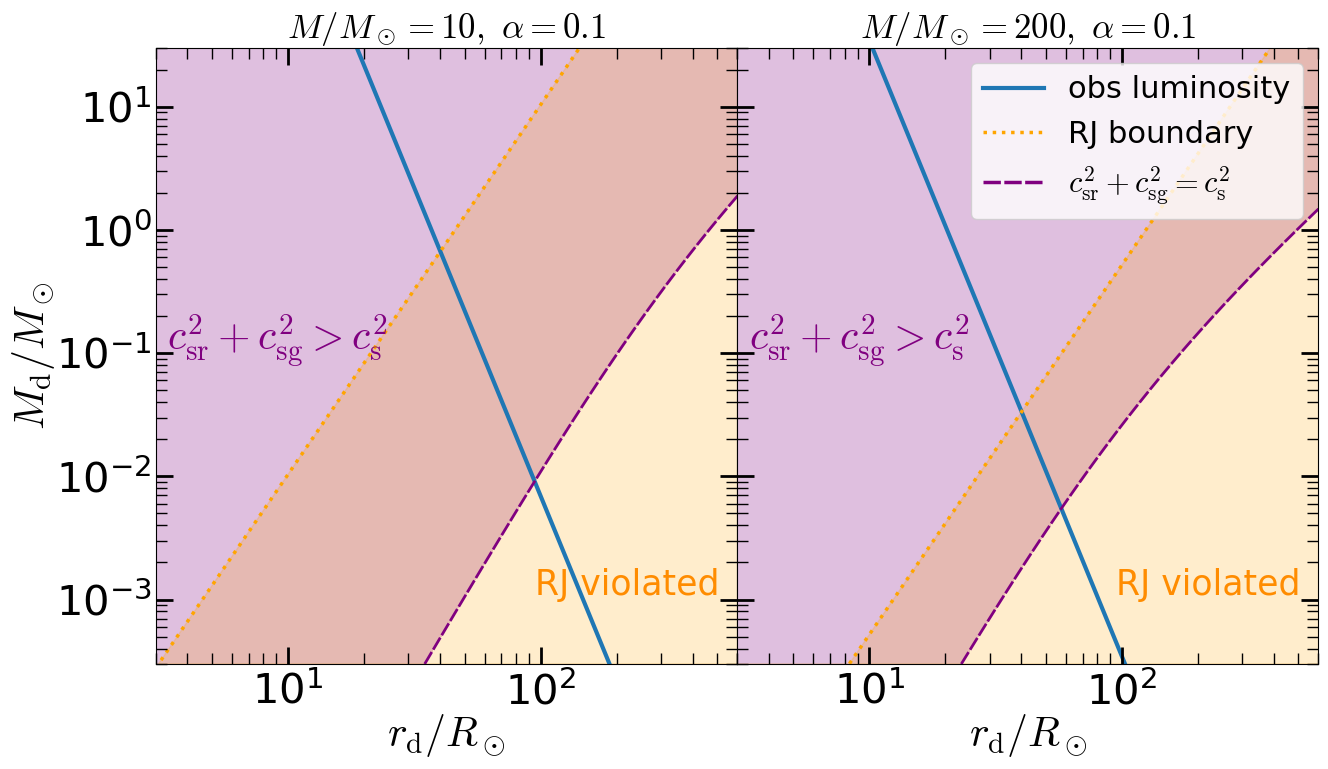}
\includegraphics[width=0.47\textwidth]{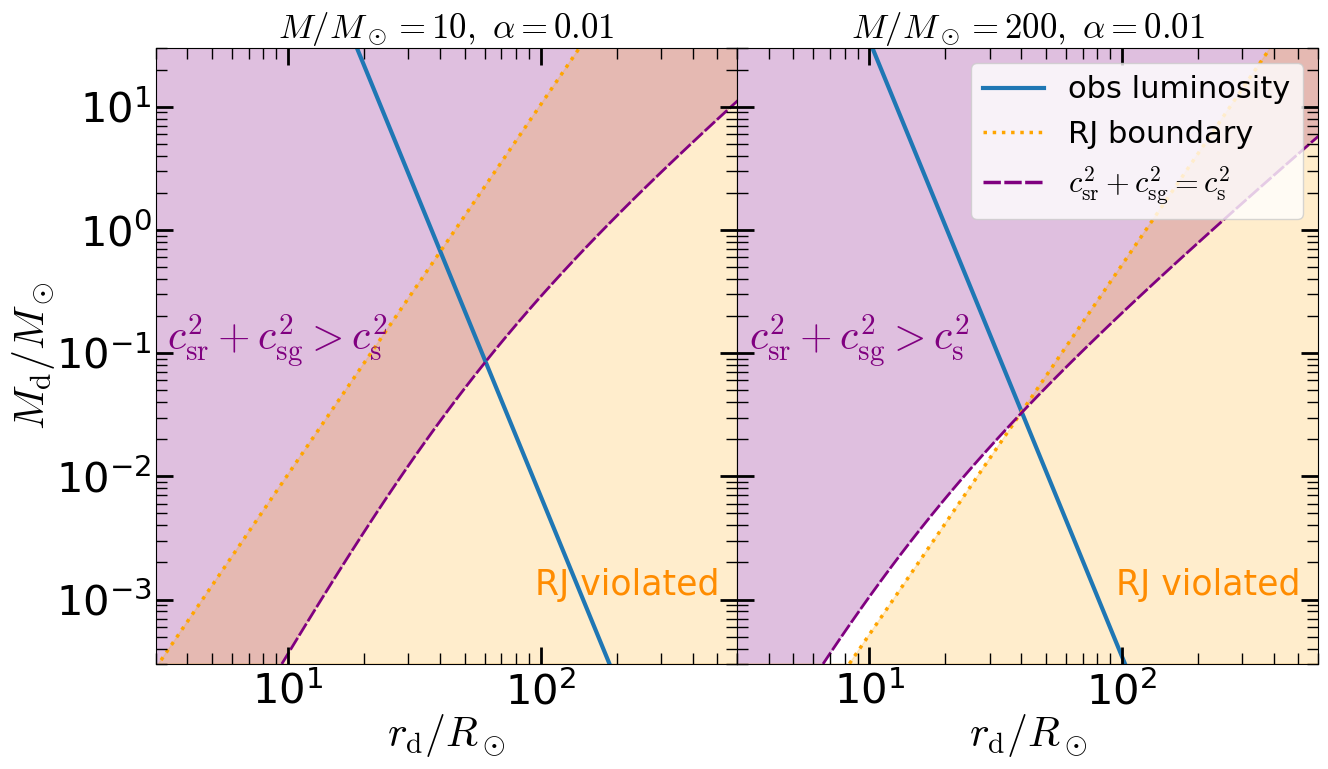}
\includegraphics[width=0.47\textwidth]{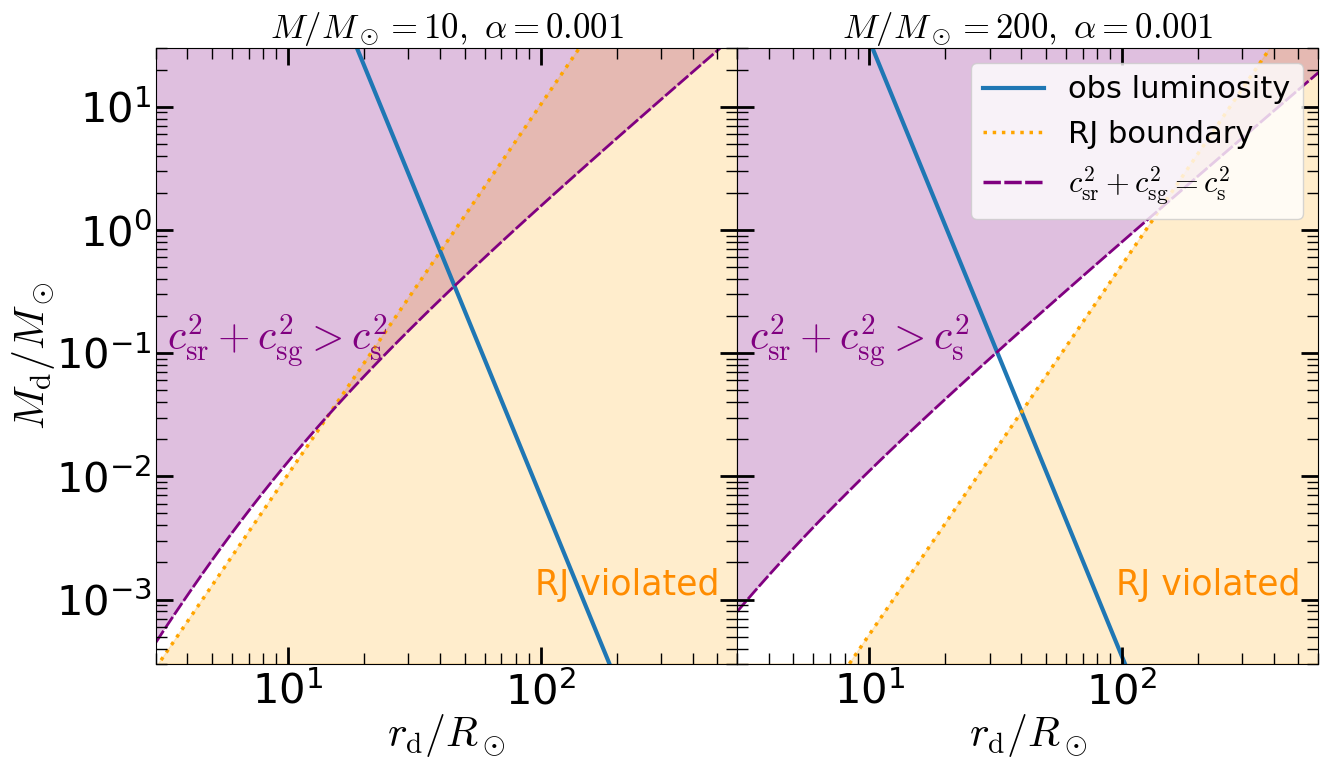}
\caption{Constraints on the outer disk accretion model for the UV/optical plateau in AT2018cow in $M_{\rm d}$-$r_{\rm d}$ parameter space, for $\alpha=0.1$ (upper), $0.01$ (middle), and $0.001$ (lower panels). The left panels show the constraints for $M=10M_\odot$ and the right panels are for $M=200M_\odot$. In each panel, we show four different constraints based on (1) observed luminosity of AT2018cow near $\nu=5.6\times 10^{14}\rm\, Hz$ at $t=1453\rm\, d$ \citep{2023ApJ...955...43C, 2025MNRAS.544L.108I} (HST F555W, blue solid line), (2) the observing frequency being on the RJ limit (orange-shaded region), and (3) self-consistency of disk pressure ($c_{\rm sr}^2 + c_{\rm sg}^2 \leq c_{\rm s}^2$, purple-shaded region). The physical solutions lie where the blue solid line passes through the unshaded region. For all panels, we fix opacity $\kappa=0.2\rm\,cm^2\,g^{-1}$ (conservative) and viscous time $t_{\rm vis} = 10\rm\,yr$.
}
\label{fig:Md_rd_constraints_main}
\end{figure}

Four constraints delimit the physically allowed region in the 
$(M_{\rm d}, \rd)$ plane, shown in Fig.~\ref{fig:Md_rd_constraints_main}
for $M = 10$ and $200\,\Msun$ and $\alpha = 0.1$, $0.01$, and $0.001$.

\begin{enumerate}

\item \textit{Observed luminosity.}
Eq.~(\ref{eq:nuLnu_RJ}), combined with $\dot{M}_{\rm d} = M_{\rm d}/t_{\rm vis}$, gives a locus $M_{\rm d} \propto \rd^{-5}$ (solid curve in 
Fig.~\ref{fig:Md_rd_constraints_main}) that must be satisfied to match  the observed $\nu L_\nu$ in the F555W band.

\item \textit{Rayleigh--Jeans condition.}
The RJ approximation is valid only if $T_{\rm eff}(\rd) > h\nu/\kB$, which requires
\begin{equation}
    \rd < 17\,\Rsun \nu_{14.75}^{-4/3}
    \Mten^{1/3}
    \lrb{\dot{M}_{\rm d}\over 0.1\,\msunyr}^{1/3}.
    \label{eq:RJ_condition}
\end{equation}
Solutions at larger $\rd$ (orange-shaded region in 
Fig.~\ref{fig:Md_rd_constraints_main}) are excluded because the 
observed bands would be inconsistent with the RJ-like SED.

\item \textit{Disk midplane pressure self-consistency.}
For the thin-disk assumption to hold, the total pressure must support the disk scale height inferred from $t_{\rm vis}$. We consider that the pressure in the outer disk midplane consists of radiation pressure $P_{\rm rad}$ and gas pressure $P_{\rm gas}$ , with their corresponding (isothermal) sound speeds 
$c_{\rm sr} \equiv (P_{\rm rad}/\rho)^{1/2}$ and $c_{\rm sg} \equiv \lrb{P_{\rm gas}/\rho}^{1/2}$, respectively. The vertical optical depth of the outer disk is
\begin{equation}
    \tau = {\kappa\Sigma(\rd)\over 2} = {3\kappa M_{\rm d}\over 8\pi \rd^2},
    \label{eq:tau}
\end{equation}
and the Eddington approximation (valid for $\tau\gg 1$) gives the midplane temperature
\begin{equation}
    T_{\rm mid}^4 = {3\over 4}\tau\, T_{\rm eff}^4(\rd),
    \label{eq:Tmid}
\end{equation}
so that $P_{\rm rad} = aT_{\rm mid}^4/3 = (a\tau/4)T_{\rm eff}^4(\rd)$ ($a$ being radiation density constant).
Requiring $c_{\rm sr} < c_{\rm s} \equiv (P/\rho)^{1/2}= (GM/\rd)^{1/2}(H/\rd)$
gives
\begin{equation}
\begin{split}
    \rd >\, &200\,\Rsun\,\alpha_{-1}^{2/7}
    \kap^{4/7} \Mten^{1/7}
    \lrb{M_{\rm d}\over\Msun}^{4/7} \lrb{t_{\rm vis}\over10\rm\,yr}^{-2/7},
\end{split}
    \label{eq:csr_constraint}
\end{equation}
where we have taken the (Rosseland-mean) opacity $\kappa = 0.2\rm\,cm^2\,g^{-1}$ as a default value ($=$ electron scattering opacity for a He-dominated or heavier composition). 

On the other hand, the gas pressure sound speed near the disk midplane is given by $c_{\rm sg} = \lrb{\kB T_{\rm mid}/\mu \mproton}^{1/2}$, where $\mu$ is the mean molecular weight and $\mproton$ is the proton mass.  Requiring $c_{\rm sg} < c_{\rm s}$ gives
\begin{equation}
\begin{split}
    \rd >\,& 350\,\Rsun\,
    \alpha_{-1}^{4/7} \kap^{1/7}\Mten^{-1/7} \times \\
    &\lrb{\mu\over4/3}^{-4/7} \lrb{t_{\rm vis}\over10\,\rm yr}^{3/7} \lrb{M_{\rm d}\over\Msun}^{2/7},
\end{split}
    \label{eq:csg_constraint}
\end{equation}
where $\alpha_{n} = \alpha/10^n$, and $\kappa=0.2\rm\,cm^2\,g^{-1}$ and $\mu=4/3$ are fiducials for a fully ionized He-dominated composition.
For a solar composition ($\kappa\simeq0.34\rm\,cm^2\,g^{-1}$, $\mu\simeq 0.61$), both the higher opacity and lower mean molecular weight will make our constraints more stringent.

Combining both radiation pressure and gas pressure, the midplane pressure self-consistency requires
\begin{equation}\label{eq:pressure_constrant_csr2_csg2_cs2}
    c_{\rm sr}^2 + c_{\rm sg}^2 \leq c_{\rm s}^2,    
\end{equation}
where the ``$<$'' sign allows for potential contribution from e.g., magnetic pressure and turbulent pressure. The region violating the above inequality (purple-shaded region in  Fig.~\ref{fig:Md_rd_constraints_main}) is unphysical because gas+radiation pressure would puff the disk to a scale height larger than inferred from the viscous timescale.

\end{enumerate}

Fig.~\ref{fig:Md_rd_constraints_main} shows that, for $\alpha \geq 10^{-3}$, the luminosity locus falls entirely within the disk-pressure-forbidden region for $M = 10\,\Msun$ or lower masses, with no allowed solutions. Increasing $M$ to $200\,\Msun$ relaxes the constraint because the stronger gravitational potential requires a smaller $M_{\rm d}$ to match the luminosity (via  eq.~\ref{eq:nuLnu_RJ}), moving the locus below the $c_{\rm sr}^2 + c_{\rm sg}^2 = c_{\rm s}^2$ boundary.

Therefore, there is a minimum compact object mass for a self-consistent thin-disk solution. For a fixed outer disk radius $\rd$ (as strongly constrained by the SED) and viscous timescale $t_{\rm vis}$ (inferred from the plateau evolution), we find that the minimum compact object mass scales as $M_{\rm min}\propto \alpha^{2/3}\kappa^{4/3}$, with the following empirical normalization (see the middle right panel in Fig. \ref{fig:Md_rd_constraints_main})
\begin{equation}\label{eq:Mmin}
    M_{\rm min} \simeq 200\,\Msun\,(\alpha/0.01)^{2/3}\kap^{4/3},
\end{equation}
For the physically motivated range $\alpha \gtrsim 0.01$ \citep{2010ApJ...713...52D, 2011ApJ...738...84H, 2020MNRAS.494.3656L}, a pure thin-disk model requires $M \gtrsim 200\,\Msun\,\kap^{4/3}$. If we consider solar composition ($\kappa \simeq 0.34\rm\,cm^2\,g^{-1}$), the constraint would be even more stringent. The minimum mass in eq. (\ref{eq:Mmin}) is in tension with the upper limit provided by the X-ray constraint $M\lesssim100\Msun$ (see \S~\ref{sec:mass_constraints}). This is consistent with the conclusion by \citet{2025arXiv251209017W}.

For an unusually lower viscosity $\alpha \sim 10^{-3}$, we find $M_{\rm min}\sim 40\Msun$. In this case, another consideration is that pair instability limits the BH mass from standard core-collapse SNe to be less than $M_{\rm PI} = 40$--$60M_\odot$ \citep[depending on metallicity, mass loss, C$(\alpha,\gamma)$O reaction rate, etc.,][]{2019ApJ...887...53F}, unless one considers rarer evolutionary channels for BH formation from very massive stars \citep{2026A&A...706A.327C}. Thus, even if $\alpha\sim 10^{-3}$ and $\kappa=0.2\rm\,cm^2\,g^{-1}$, there is only a narrow window of $40M_\odot \lesssim M\lesssim M_{\rm PI}$ that satisfies all the basic constraints. Therefore, we consider the accretion powered thin-disk interpretation to be moderately fine-tuned.





\subsection{Super-Eddington wind and outer disk irradiation}
\label{sec:wind_model}

\begin{figure}
\centering
\includegraphics[width=0.4\textwidth]{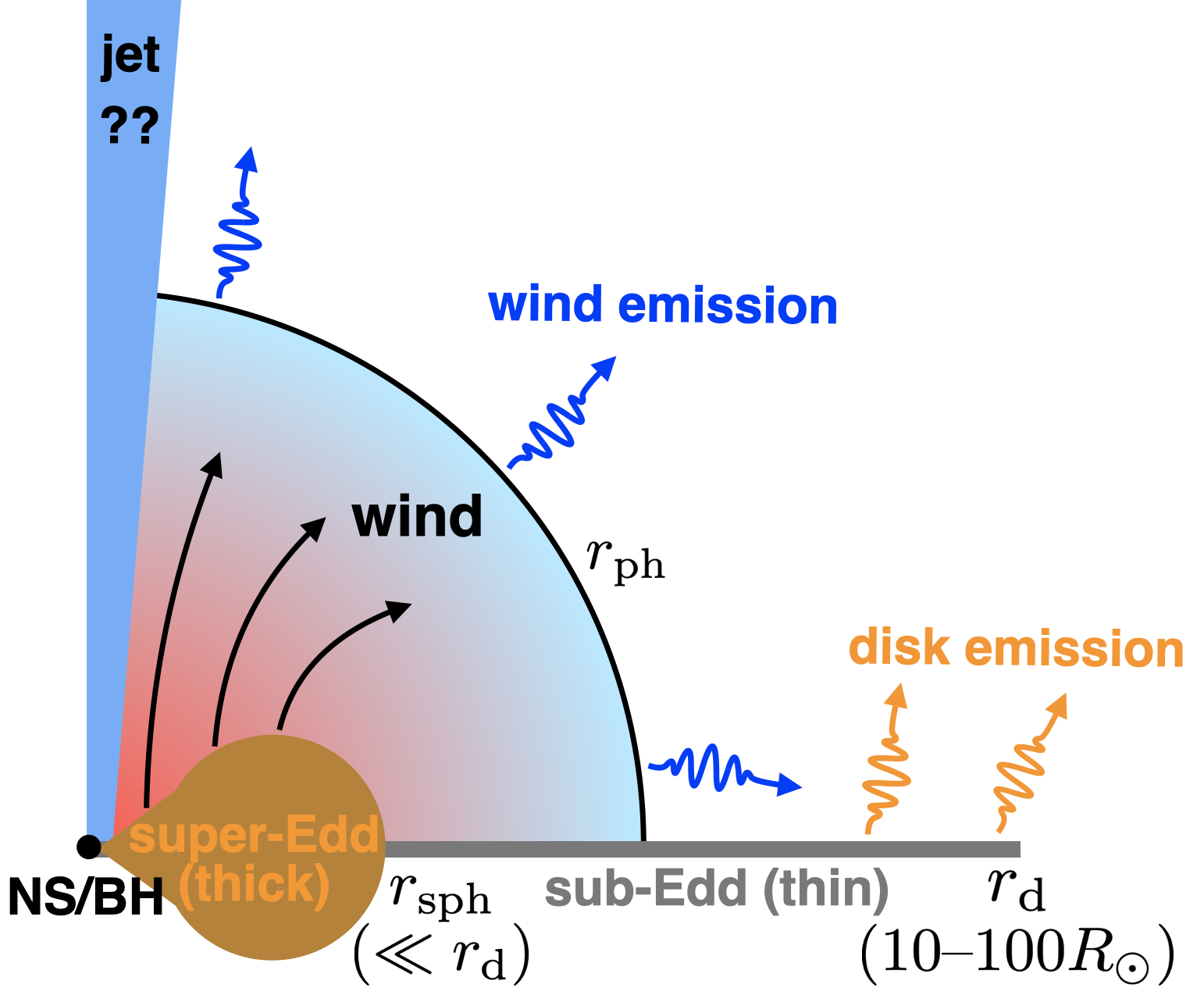}
\caption{Schematic picture for the disk+wind+irradiation model for the late-time UV plateau. The inner, super-Eddington region of the accretion disk below the spherization radius ($r < r_{\rm sph}$) is geometrically thick, which leads to a radiatively driven, optically thick wind. The photospheric radius of the wind is denoted as $r_{\rm ph}$. The outer, sub-Eddington region of the accretion disk at $r>r_{\rm sph}$ is geometrically thin. The disk outer radius is denoted as $r_{\rm d}$. The scenario shown here is for $r_{\rm ph} < r_{\rm d}$, so the disk surface emission from $r_{\rm ph} < r < r_{\rm d}$ is directly observable, but $r_{\rm ph} > r_{\rm d}$ (not shown here) is also possible, in which case the disk surface emission is not directly observable.
}
\label{fig:sketch}
\end{figure}

We now consider a new disk model that includes the effects of a super-Eddington wind, as schematically shown in Fig. \ref{fig:sketch}.

Our model is motivated by the fact that the global accretion rate generally exceeds the Eddington value,
\begin{equation}
    \dot{M}_{\rm d} > \dot{M}_{\rm Edd} \equiv {10 L_{\rm Edd}\over c^2},
\end{equation}
where the Eddington luminosity is $L_{\rm Edd} = 4\pi GMc/\kappa = 2.5\times10^{39}\mr{\, erg\,s^{-1}}
\Mten \kap^{-1}$. In the super-Eddington regime, the inner disk is geometrically thick and is bounded by the
spherization radius \citep{1973A&A....24..337S}
\begin{equation}
    r_{\rm sph} = {\kappa\dot{M}_{\rm d}\over 4\pi c}
    = {GM\dot{M}_{\rm d}/ L_{\rm Edd}}.
    \label{eq:rsph}
\end{equation}
At $r < r_{\rm sph}$ the disk is radiation pressure dominated and launches a
powerful wind \citep[e.g.,][]{1999MNRAS.310.1002S, 1999MNRAS.303L...1B, 2014ARA&A..52..529Y},
causing the accretion rate to decrease inward as
\begin{equation}
    \dot{M}(r) = \dot{M}_{\rm d}\lrb{r\over r_{\rm sph}}^s, \quad r < r_{\rm sph},
    \label{eq:Mdot_r}
\end{equation}
with $s \simeq 0.5$ from recent simulations \citep{2024ApJ...973..141G}.
At $r > r_{\rm sph}$ the disk is geometrically thin and the accretion rate is
constant at $\dot{M}_{\rm d}$.

The wind launched from radius $r < r_{\rm sph}$ has an asymptotic velocity comparable to the 
local Keplerian speed, $v(r) \simeq \sqrt{GM/r}$ \citep{2016MNRAS.458.1214Q, 2016MNRAS.459..171S}.
The wind optical depth near the launching point satisfies
\begin{equation}
    {\tau(r)\,v(r)\over c} \simeq
    {\kappa\dot{M}(r)\over 4\pi r c} = \lrb{r\over r_{\rm sph}}^{-1/2} > 1,
    \label{eq:trapping}
\end{equation}
so radiation is advectively trapped near the launching point for any $r < r_{\rm sph}$.
Accounting for adiabatic losses during wind expansion, the radiative power 
escaping from the wind launched near radius $r$ is
\begin{equation}
    {\d L_{\rm rad}\over \d \ln r} \simeq (\tau v/c)^{-2/3} L_{\rm acc}
\simeq (r/r_{\rm sph})^{-1/6} L_{\rm Edd},
\end{equation}
where $L_{\rm acc}(r) = GM\dot{M}(r)/r$ is the characteristic accretion power near radius $r$.
Integrating from the inner edge $r_{\rm in} \simeq 10r_{\rm g}$ (appropriate for
both NSs and weakly spinning BHs) to $r_{\rm sph}$ and 
taking the limit $r_{\rm sph} \gg r_{\rm in}$ gives the total wind radiative power
\begin{equation}\label{eq:Lrad}
\begin{split}
    &L_{\rm rad,tot} = \int_{r_{\rm in}}^{r_{\rm sph}} {\d L_{\rm rad}\over \d \ln r}\d \ln r \simeq 6\lrb{\dot{M}_{\rm d}\over\dot{M}_{\rm Edd}}^{1/6}
    L_{\rm Edd}\\
    &\simeq 5.4\times10^{40}\mr{\, erg\, s^{-1}}
    \kap^{-5/6}\Mten^{5/6} \lrb{\dot{M}_{\rm d}\over10^{-3}\msunyr}^{1/6}.
\end{split}
\end{equation}
The weak dependence on $\dot{M}_{\rm d}$ means that a compact object of
mass $M \sim 10\,\Msun$ accreting at $\dot{M}_{\rm d} \sim 10^{-3}\,\msunyr$
can power $L_{\rm rad,tot} \sim 10^{40}\rm\, erg\, s^{-1}$, broadly
consistent with the UV/optical plateau luminosity of AT2018cow.

The wind photosphere is set by mass loss near $r_{\rm sph}$, where the
mass flux is largest and the outflow velocity is slowest.
Taking the entire $\dot{M}_{\rm d}$ to be ejected at $v(r_{\rm sph})$, the
photospheric radius is
\begin{equation}\label{eq:rph}
\begin{split}
     r_{\rm ph} &\simeq {\kappa\dot{M}_{\rm d}\over 4\pi v(r_{\rm sph})}
    = {\kappa\dot{M}_{\rm d}^{3/2}\over 4\pi L_{\rm Edd}^{1/2}} \simeq 72\,\Rsun\,
    \kap^{3/2}\times \\
    & \lrb{\dot{M}_{\rm d}\over10^{-3}\msunyr}^{3/2}
    \Mten^{-1/2}.
\end{split}
\end{equation}
At high accretion rates the photosphere may be so extended that the
effective temperature at $r_{\rm ph}$ falls below the hydrogen recombination
temperature $T_{\rm rec} \approx 6000\rm\, K$. We therefore adopt a photospheric radius
$r_{\rm ph} = \min\left[(L_{\rm rad,tot}/4\pi\sigma_{\rm SB}T_{\rm rec}^4)^{1/2},\,
\kappa\dot{M}_{\rm d}^{3/2}/4\pi L_{\rm Edd}^{1/2}\right]$. In practice, a photospheric temperature near $T_{\rm rec}$ does not fit the UV/optical plateau SED of AT2018cow well, so this is simply a physics guard that has no practical importance.

If radiation thermalizes efficiently near $r_{\rm ph}$, the color temperature of the wind emission is
\begin{equation}
\begin{split}
    T_{\rm w}' =\, &\lrb{L_{\rm rad,tot}\over4\pi r_{\rm ph}^2\sigma_{\rm SB}}^{1/4}\simeq 4.2\times10^4\mr{\, K\,}
    \kap^{-23/24} \times\\
    &\Mten^{11/24} \lrb{\dot{M}_{\rm d}\over10^{-3}\msunyr}^{-17/24}.
\end{split}
    \label{eq:Tw}
\end{equation}
In practice, the opacity near the photosphere is most likely scattering dominated, so the corresponding thermalization is inefficient and the observed color temperature is somewhat higher than $T_{\rm w}'$. We parameterize this with a color correction factor,
\begin{equation}
    T_{\rm w} = f_{\rm col}T_{\rm w}',
\end{equation}
where $f_{\rm col} \geq 1$ is a free parameter discussed below.

Let us denote the ratio of Planck-mean \textit{absorption} opacity to Rosseland-mean \textit{total} opacity as $f_{\rm abs} = \kappa_{\rm a}/\kappa$. We assume for simplicity that $f_{\rm abs}$ is independent of radius and the density profile is given by $\rho\propto r^{-2}$ (constant mass-loss rate), since the effective optical depth $\tau_{\rm eff}\simeq \sqrt{\tau_{\rm a}\tau} = f_{\rm abs}^{1/2}\tau$, we obtain the thermalization radius (where $\tau_{\rm eff}\simeq 1$)
\begin{equation}
    r_{\rm th} \simeq f_{\rm abs}^{1/2} r_{\rm ph},
\end{equation}
and the corresponding Rosseland-mean optical depth is given by $\tau(r_{\rm th}) \simeq f_{\rm abs}^{-1/2}$. The color temperature of the wind emission is set by the radiation temperature at the thermalization radius, i.e., $T_{\rm w} = T_{\rm th} \equiv T(r_{\rm th})$.

Since the diffusive luminosity at the thermalization radius is roughly given by
\begin{equation}
    L_{\rm bol} \simeq 4\pi r_{\rm th}^2 {a T_{\rm th}^4 c\over \tau(r_{\rm th})} \simeq 4\pi r_{\rm ph}^2 {f_{\rm abs}^{3/2}} {acT_{\rm th}^4},
\end{equation}
and $L_{\rm bol} = 4\pi r_{\rm ph}^2 \sigma_{\rm SB}(T_{\rm w}')^4$, we obtain the radiation temperature at the thermalization radius
\begin{equation}
    T_{\rm th} = f_{\rm col} T_{\rm w}', \ \ f_{\rm col}\simeq {1\over 4^{1/4}f_{\rm abs}^{3/8}}.
\end{equation}
The weak dependence of the color correction factor $f_{\rm col}$ on $f_{\rm abs}$ shows that $f_{\rm col}$ is unlikely more than 10, as that would require $f_{\rm abs}\lesssim 10^{-3}$. Another physical limit for the color correction factor is set by the temperature near the spherization radius, where we expect thermalization to hold; this gives an upper limit
\begin{equation}
\begin{split}
f_{\rm col,sph} &\simeq \lrb{r_{\rm ph}\over r_{\rm sph}}^{3/4} = \lrb{\dot{M}_{\rm d} c^2\over L_{\rm Edd}}^{3/8}\\
&= 43\, \kap^{3/8} \Mten^{-3/8} \lrb{\dot{M}_{\rm d}\over 10^{-3}\,\msunyr}^{3/8}.
\end{split}
\end{equation}
This is less constraining than that from $f_{\rm abs}$, except for the most massive accretors $M\gg 10^2M_\odot$. For the above reasons, we take a flat prior for $\log f_{\rm col}$ in the range of [0, 1]. 



Modeling the detailed radiative transfer and thermalization in the wind is left for future work. We consider the emission from the wind as a blackbody at temperature $T_{\rm w}$,
normalized to the total luminosity $L_{\rm rad,tot}$,
\begin{equation}
    L_{\rm w,\nu} = {\pi B_\nu(T_{\rm w})\over\sigma_{\rm SB}T_{\rm w}^4}
    \,L_{\rm rad,tot}.
    \label{eq:Lw}
\end{equation}

\begin{figure*}
\centering
\includegraphics[width=0.8\textwidth]{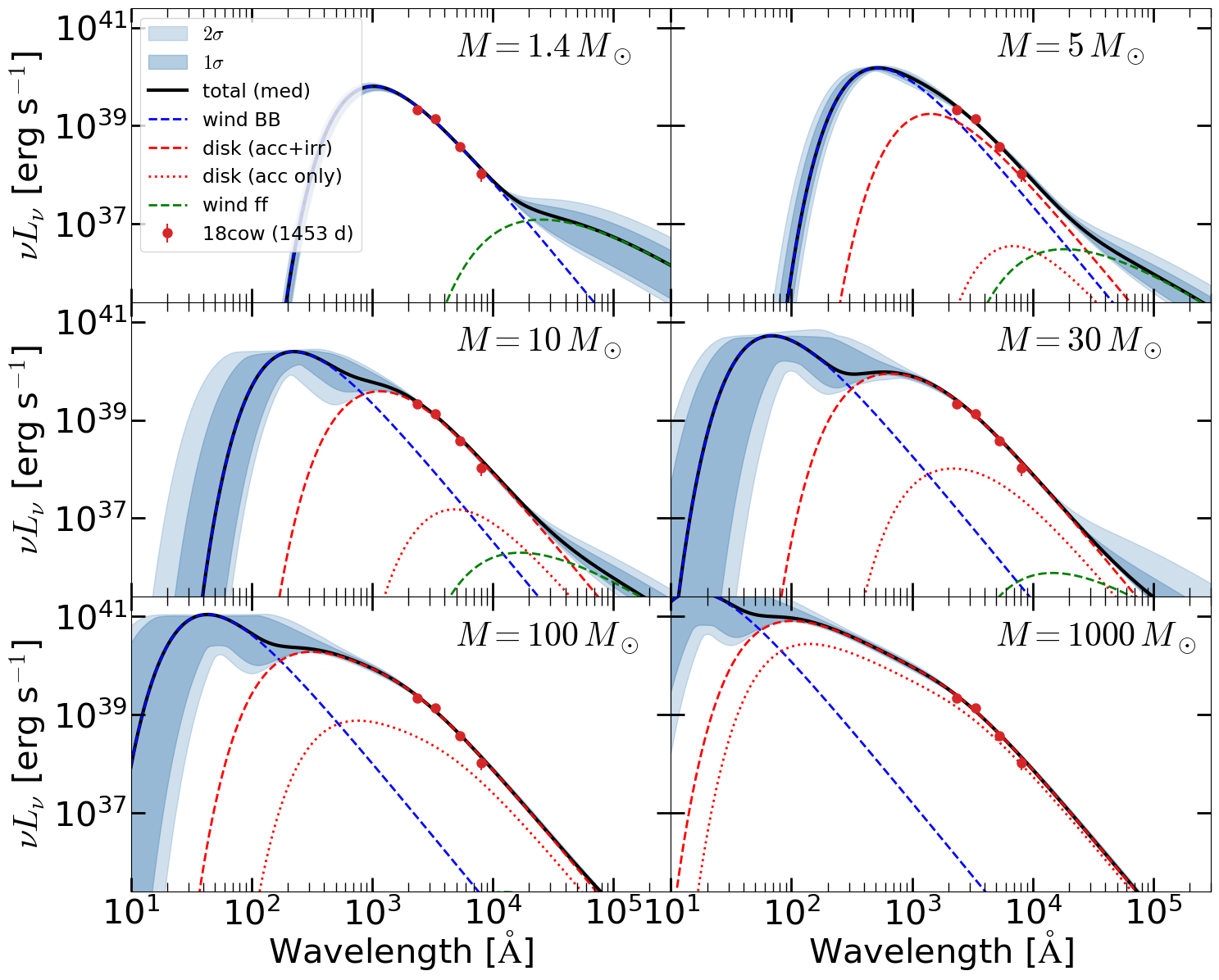}
\caption{Monte Carlo fit to the AT2018cow data at $t=1453\rm\, d$ for different compact object masses $M$. The blue bands show the $1\sigma$ and $2\sigma$ posteriors evaluated at each frequency. The curves show the predictions for a median set of parameters in each panel: the black solid line shows the total, and the colored dashed lines show the contributions from different emission components. For $M\lesssim 5M_\odot$, the UV/optical emission is dominated by emission from the super-Eddington wind launched from the inner disk; for $10\lesssim M \lesssim 100M_\odot$, outer disk irradiation by the wind emission dominates the UV/optical (direct accretion power is subdominant) and, for increasing $M$, the wind emission shifts to the extreme-UV and X-ray bands; for $M\gtrsim 300M_\odot$, direct accretion power from the outer disk plays a dominant role, and the wind emission entirely shifts into the X-ray band.
}
\label{fig:sed_per_mass}
\end{figure*}

\subsubsection{Irradiation and emission from the outer disk}

When $r_{\rm ph} < \rd$, the geometrically thin outer disk at $r_{\rm ph} < r < \rd$ receives
both viscous accretion heating and irradiation by the wind photosphere.
The accretion contribution to the effective surface temperature (hereafter denoted as $T_{\rm eff,acc}$) is given by eq.~(\ref{eq:Teff}), and the irradiation contribution is
\begin{equation}\label{eq:Teffirr}
    T_{\rm eff, irr}^4(r) = \lrb{T_{\rm w}'}^4\, f_{\rm irr}(r/r_{\rm ph}),
\end{equation}
where $f_{\rm irr}(x)$ is the geometric irradiation factor for a spherical
photosphere illuminating a thin equatorial disk \citep{2023MNRAS.519.1409L}:
\begin{equation}
    f_{\rm irr}(x) = {2\over\pi}\int_{x^{-1}}^{1}\d\mu\,
    {(x\mu - 1)\sqrt{1-\mu^2}\over(x^2+1-2x\mu)^2}.
    \label{eq:firr}
\end{equation}
The total effective temperature is $T_{\rm eff}^4(r) = T_{\rm eff,acc}^4(r)
+ T_{\rm eff,irr}^4(r)$, and the outer disk spectral luminosity is
\begin{equation}
    L_{\rm d,\nu} = 4\pi^2\int_{r_{\rm ph}}^{\rd} B_\nu(T_{\rm eff}(r))\,r\,{\rm d}r.
    \label{eq:Ldisk}
\end{equation}
Note that, in the irradiation calculation (eq. \ref{eq:Teffirr}), we use $T_{\rm w}'$, rather than the color temperature $T_{\rm w}$, because the disk irradiation is determined by the bolometric radiation energy flux, while $f_{\rm col}$ only redistributes the emergent wind spectrum in frequency.
Lastly, any disk emission from $r < r_{\rm ph}$ is neglected as it is overwhelmed by the wind photosphere.

\subsubsection{Free-free emission}

The ionized disk wind emits free-free radiation that dominates the SED at NIR-MIR wavelengths. Let us take the wind density profile as $\rho(r) = A_{\rm w}/r^2$ with $A_{\rm w} = \dot{M}_{\rm d}/[4\pi v(r_{\rm sph})]$. The free-free absorption opacity at frequency $\nu$ is
\begin{equation}
    \kappa_{{\rm ff},\nu} = \frac{Z^2Q\,n_{\rm e}}{AT_{\rm e}^{1/2}\,\nu^3}
    \lrb{1 - \mr{e}^{-h\nu/\kB T_{\rm e}}},
    \label{eq:kff}
\end{equation}
where $n_{\rm e} = \rho/(\mu_{\rm e} m_p)$ with the mean molecular weight per electron $\mu_{\rm e}$ ($=2$ as a fiducial value), $T_{\rm e}$ is the electron temperature, $A$ and $Z$ are the mass and charge numbers of the dominating species, $Q = 2.21\times10^{32}$ in CGS units, and we have taken the Gaunt factor $g_{\rm ff} = 1$ (see below). For any composition dominated by H or He, we set $Z^2/A=1$ throughout. We take a fiducial electron temperature of $T_{\rm e} = 2\times 10^4\rm\,K$, which only weakly affects our results as the free-free emission in the Rayleigh-Jeans limit is independent of $T_{\rm e}$ (see below).

Free-free photons emerge from the thermalization radius $r_{\rm th,\nu}$ where
$\tau_{\rm eff}(r_{\rm th}) \simeq 1$, with
\begin{equation}
    \tau_{\rm eff}(r) = \sqrt{\tau_{{\rm ff},\nu}(r)\,
    \bigl[\tau_{{\rm ff},\nu}(r) + \tau_{\rm es}(r)\bigr]},
    \label{eq:taueff_ff}
\end{equation}
accounting for both free-free absorption and electron scattering.
Writing $\kappa_{{\rm ff},\nu} \equiv C_\nu/r^2$ with
$C_\nu = Q A_{\rm w} (1-e^{-h\nu/\kB T_{\rm e}})/(\mu_{\rm e} m_p T_{\rm e}^{1/2}\nu^3)$,
the condition $\tau_{\rm eff}(r_{\rm th}) = 1$ reduces to the following cubic equation
\begin{equation}
    w^3 - \frac{A_{\rm w}^2 C_\nu \kappa}{3}\,w - \frac{A_{\rm w}^2 C_\nu^2}{9} = 0,
    \quad w \equiv r_{\rm th,\nu}^2.
    \label{eq:ff_cubic}
\end{equation}
Applying Kirchhoff's law ($j_\nu = \rho\kappa_{{\rm ff},\nu}B_\nu$) and
integrating over $r > r_{\rm th}$ (where $j_\nu\propto r^{-4}$), the
spectral luminosity is given by
\begin{equation}
    L_{{\rm ff},\nu} = \frac{32\pi^2 h Q}{c^2 \mu_{\rm e} m_p}
    \frac{A_{\rm w}^2}{r_{\rm th,\nu}}\,T_{\rm e}^{-1/2}\,e^{-h\nu/\kB T_{\rm e}},
    \label{eq:Lff_general}
\end{equation}
where the exponential factor follows from the identity $(1-e^{-x})/(e^x-1) = e^{-x}$ applied to the product $\kappa_{{\rm ff},\nu}\,B_\nu$. When $r_{\rm th,\nu} < r_{\rm ph}$, we further multiply eq.~(\ref{eq:Lff_general}) by $e^{-r_{\rm ph}/r_{\rm th,\nu}}$ to suppress emission absorbed within the wind photosphere.

In the Rayleigh--Jeans limit ($h\nu \ll \kB T_{\rm e}$) and when electron
scattering is sub-dominant near the thermalization radius,
the cubic~(\ref{eq:ff_cubic}) reduces to $w^3 = A_{\rm w}^2C_\nu^2/9$, giving
$r_{\rm th,\nu} \propto A_{\rm w}^{2/3}\nu^{-2/3}$.
Substituting into eq.~(\ref{eq:Lff_general}), the $T_{\rm e}$ dependences
cancel and one recovers the classic wind free-free spectrum
\citep{1975MNRAS.170...41W}
\begin{equation}\label{eq:Lff}
\begin{split}
    \nu L_{\rm ff,\nu}&(\mbox{RJ limit}) \simeq 2.7\times10^{37}\rm\, erg\, s^{-1}\,
    \nu_{14}^{5/3}\times\\
    &\kap^{2/3}(\mu_{\rm e}/2)^{-2/3}\Mten^{-2/3}\lrb{\dot{M}_{\rm d}\over10^{-3}\msunyr}^{2},
\end{split}
\end{equation}
where $\nu_{14}\equiv\nu/10^{14}\rm\,Hz$.
The scalings follow from $A_{\rm w}\propto\dot{M}_{\rm d}^{3/2}(\kappa/M)^{1/2}$,
giving $A_{\rm w}^{4/3}\propto\dot{M}_{\rm d}^2 M^{-2/3}\kappa^{2/3}$; the $T_{\rm e}$
independence holds because the factors of $T_{\rm e}$ from $C_\nu\propto T_{\rm e}^{-3/2}$
and from $T_{\rm e}^{-1/2}$ in eq.~(\ref{eq:Lff_general}) exactly cancel. Because $\nu L_{\rm ff,\nu} \propto \dot{M}_{\rm d}^2 M^{-2/3}$ and the
posterior for $\dot{M}_{\rm d}$ shifts to larger values at lower $M$
(to match the observed UV luminosity), the free-free flux at $\lambda \sim 1$--$10\,\mu$m
is significantly brighter for lower-mass accretors.

Finally, the Gaunt factor $g_{\rm ff}$ enters as $Q\propto g_{\rm ff}$, so
eq.~(\ref{eq:Lff}) scales as $g_{\rm ff}^{2/3}$.
For NIR-MIR frequencies and typical electron temperatures $T_{\rm e}\sim\mr{few}\times 10^4\rm\,K$ ($h\nu/\kB T_{\rm e}\sim0.05$--$0.5$), the Gaunt factor is given by \citep{2011piim.book.....D}
\begin{equation}
\begin{split}
    g_{\rm ff} &\approx \ln \lrsb{\mr{e} + \lrb{\lrb{\kB T_{\rm e}}^{3/2}\over h\nu \sqrt{Z^2\mr{Ry}}}^{\sqrt{3}/\pi}}\\
    &\approx \ln \lrsb{\mr{e} + \lrb{0.52\,Z^{-1} T_{\rm e,4}^{3/2}\nu_{14}^{-1}}^{\sqrt{3}/\pi}},
\end{split}
\end{equation}
where $\mr{Ry}=13.6\rm\,eV$ is the Rydberg energy. We find $1\lesssim g_{\rm ff}\lesssim 1.8$ and the correction factor of $g_{\rm ff}^{2/3}$ is modest; this can be absorbed into the model uncertainty, justifying our adoption of $g_{\rm ff}\simeq 1$.

The full model SED is
\begin{equation}
    L_\nu = L_{\rm w,\nu} + L_{\rm d,\nu} + L_{\rm ff,\nu}.
    \label{eq:Ltot}
\end{equation}
For fixed $M$, the model has three free parameters:
$\log_{10}(\rd/\Rsun)$, $\log_{10}(\dot{M}_{\rm d}/\msunyr)$,
and $\log_{10}f_{\rm col}$. Note that the disk+wind model is agnostic about the viscous evolution of the disk --- once the posteriors for $\rd, \dot{M}_{\rm d}$ are obtained, we can then infer the current disk mass $M_{\rm d}\simeq \dot{M}_{\rm d}t_{\rm vis}$ by \textit{assuming} a viscous timescale $t_{\rm vis}$. We compare the posterior from the SED fits with the expectation from the disk evolution in the aftermath of (micro-)tidal disruption events in \S \ref{sec:micro-TDEs}.

\section{SED fits and posteriors}\label{sec:SED_fits}

\begin{figure}
\centering
\includegraphics[width=0.48\textwidth]{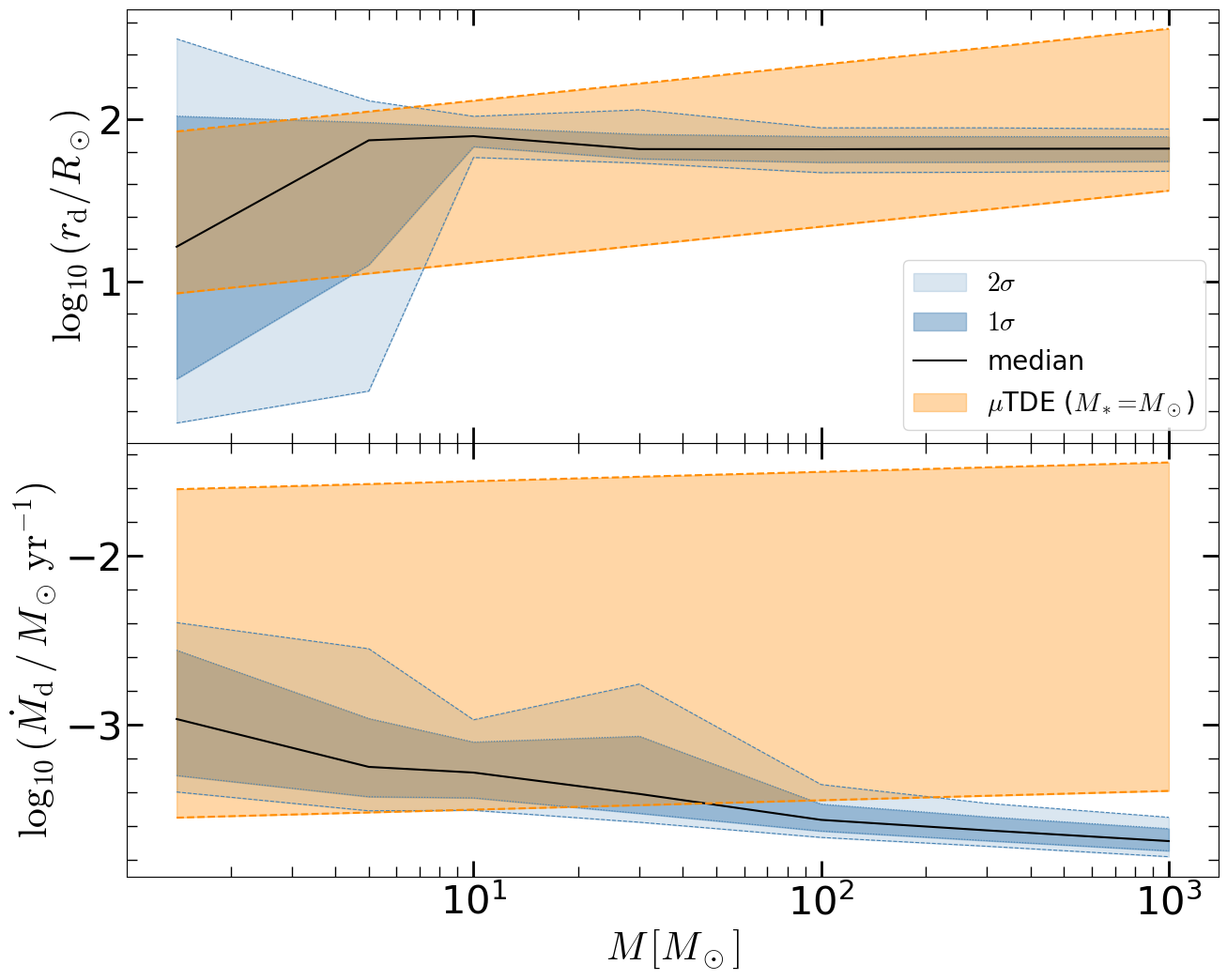}
\caption{Marginalized posteriors for $r_{\rm d}$ and $\dot{M}$ for different compact object masses. The orange band in each panel shows the expectations from disk evolution after a micro-TDE of a star with $M_*=M_\odot$ and $R_*=R_\odot$ (as an example), with disk parameters $\alpha h_{\rm thick}^2\in[10^{-3},\,0.1]$, $t_{\rm vis,thin}\in[5,\,30]\,\rm yr$, and $f_{\rm j}\in [0.1, 1]$ (see \S \ref{sec:micro-TDEs} for details).
}
\label{fig:posterior_rout_mdot}
\end{figure}

\begin{figure*}
\centering
\includegraphics[width=0.95\textwidth]{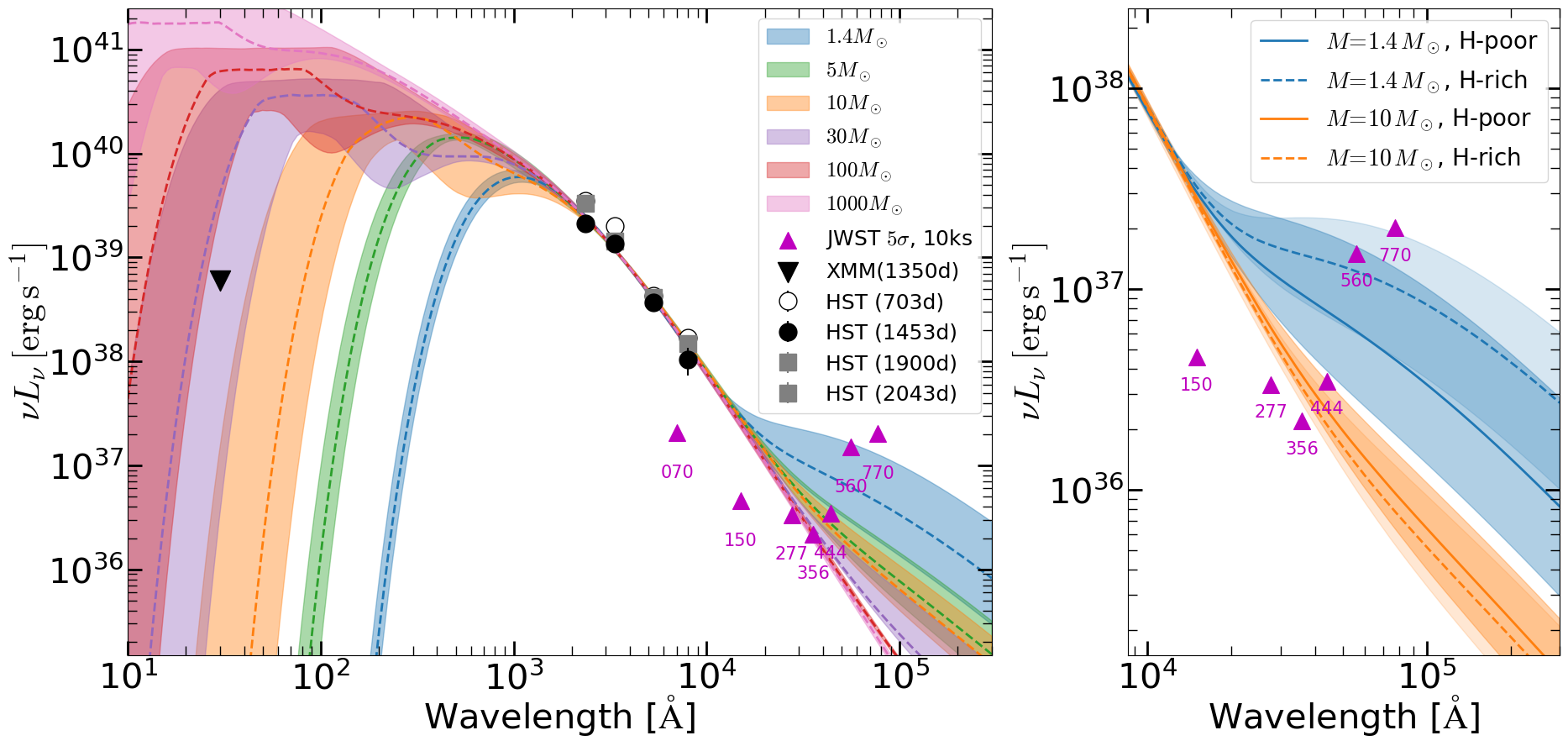}
\caption{\textbf{Left:} Posterior SED bands (16th--84th percentile, shaded; median, dashed) for
compact object masses $M \in \{1.4,\,5,\,10,\,30,\,100,\,1000\}\,\Msun$ from X-ray to MIR. Grey/black circles/squares show HST photometry at several late epochs for reference; only the t=1453 d points are used in the fits. Black triangle (downward): XMM-Newton upper limit at $t\simeq 1350\rm\, d$. Magenta triangles (upward): JWST $5\sigma$ point-source limits for a $10\,\rm ks$ exposure (filter names labeled). All models fit the UV/optical data equally well, but the NIR-MIR emission (dominated by wind free-free) drops with increasing accretor mass, providing a test with JWST. \textbf{Right:} Posterior SED bands (16th--84th percentile) in the NIR-MIR for $M=1.4$ and $10\Msun$ and for two different disk compositions: solar $=$ H-rich, and He-dominated $=$ H-poor. The solid and dashed lines show the median in each case. 
}
\label{fig:sed_comparison}
\end{figure*}

We fit the model (eq.~\ref{eq:Ltot}) to the AT2018cow $t = 1453\rm\,d$ HST
photometry in four bands (corrected for Galactic extinction): F225W = 24.93 $\pm$ 0.11, F336W = 25.03 $\pm$ 0.07, F555W = 25.94 $\pm$ 0.12, F814W = 26.87 $\pm$ 0.33 mag (AB) \citep{2023ApJ...955...43C}. The UV/optical plateau is slowly evolving (see Fig. \ref{fig:sed_comparison}), so we expect the fitting results to be similar for other epochs. For each fixed compact object mass $M \in \{1.4,\,5,\,10,\,30,\,100,\,300,\,1000\}\,\Msun$, we explore the three-dimensional parameter space $(\log_{10}(\rd/\Rsun),\,\log_{10}(\dot{M}_{\rm d}/\msunyr),\,\log_{10}f_{\rm col})$ using the dynamic nested sampler \texttt{dynesty} \citep{2020MNRAS.493.3132S} with uniform priors:
\begin{equation}
\begin{aligned}
    &\log_{10}(\rd/\Rsun) \in [0,\, 3], \\
    &\log_{10}(\dot{M}_{\rm d}/\msunyr) \in [-5,\, 0], \\
    &\log_{10}f_{\rm col} \in [0,\, 1].
\end{aligned}
\label{eq:priors}
\end{equation}
The likelihood is taken to be Gaussian in $F_\nu$ space:
\begin{equation}
    \ln\mathcal{L} = -{1\over2}\sum_{i}
    \lrb{F_{\nu,i}^{\rm model} - F_{\nu,i}^{\rm obs}\over\sigma_{\nu,i}}^2,
    \label{eq:likelihood}
\end{equation}
where $F_\nu^{\rm model} = L_\nu/(4\pi D_{\rm L}^2)$ and $D_{\rm L} = 60\rm\,Mpc$.

Fig.~\ref{fig:sed_per_mass} shows the posterior SED bands ($1\sigma$ and $2\sigma$)
for each mass, together with the HST data points.
All tested masses from $1.4\,\Msun$ (NS) to $1000\,\Msun$
produce acceptable fits to the UV/optical SED, and the log-evidences $\ln\mathcal{Z}$
differ by at most $\simeq 2.6$ across the full range, corresponding to odds of $\mr{e}^{2.6}\simeq 13$ --- inconclusive according to the Jeffreys scale.
The four-band HST photometry alone cannot distinguish between accretor masses.

The dominant emission component varies systematically with the accretor mass $M$:
\begin{itemize}
    \item For $M\lesssim 5\,\Msun$, the wind photosphere dominates the UV/optical;
    \item For $10\lesssim M\lesssim 100\,\Msun$, the outer disk irradiated by the wind
provides the bulk of the emission in the UV/optical whereas direct accretion power is subdominant;
    \item For $M\gtrsim 300\,\Msun$, thin-disk accretion power dominates the UV/optical and the wind
photosphere shifts to the EUV/X-ray bands.
\end{itemize}

Fig.~\ref{fig:posterior_rout_mdot} shows the marginalized posteriors of
$r_{\rm d}$ and $\dot{M}_{\rm d}$, as summarized below:
\begin{itemize}
    \item For $M\geq 10\,\Msun$, the outer disk radius converges to
$\rd \simeq 60$--$90\,\Rsun$ (68\% credible interval), and the corresponding emitting area is roughly consistent with the blackbody-equivalent radius $R_{\rm BB}\sim 40\,\Rsun$ inferred from the observed optical luminosity (eq.~\ref{eq:R_BB}). At these masses the accretion rate is constrained to $\dot{M}_{\rm d}\simeq (2\text{--}10)\times10^{-4}\,\msunyr$, decreasing weakly
with $M$ as the dominant UV channel transitions from the irradiated disk
emission to viscously powered disk emission.
\item For $M\leq 5\,\Msun$ the posterior on $\rd$ broadens significantly because the wind photosphere, which dominates the UV/optical, depends only weakly on $\rd$; at the same time $\dot{M}_{\rm d}$ shifts to higher values ($\sim10^{-3}\,\msunyr$) to maintain the observed luminosity. 
\end{itemize}
In \S \ref{sec:micro-TDEs}, we show that $r_{\rm d}$ and $\dot{M}_{\rm d}$ we infer are broadly consistent with the expected disk evolution from micro-TDEs (orange bands). The color correction factor is only weakly constrained to $f_{\rm col}\simeq1.6$--$3.7$ for all masses. This is consistent with scattering-dominated wind photospheres where $f_{\rm col}\simeq4^{-1/4}f_{\rm abs}^{-3/8}$ and $f_{\rm abs} \ll 1$.

Fig.~\ref{fig:sed_comparison} shows the posterior SED bands for all tested masses extending from the X-ray band to $\lambda \sim 10\,\mu\rm m$, together
with the HST data and the JWST 5$\sigma$ point-source sensitivity for a $10\,\rm ks$
integration. At UV/optical wavelengths all models are consistent with the HST data, as expected
from the mass degeneracy discussed above. The models diverge at NIR-MIR wavelengths ($\lambda \gtrsim 1\,\mu\rm m$), where the free-free emission dominates and scales as
$\nu L_{\rm ff,\nu} \propto \dot{M}_{\rm d}^2 M^{-2/3}$
(eq.~\ref{eq:Lff}). Because the posterior accretion rate shifts to higher values at lower $M$
(from $\dot{M}_{\rm d}\sim 2\times10^{-4}\,\msunyr$ at $M=1000\,\Msun$ to
$\sim10^{-3}\,\msunyr$ at $M=1.4\,\Msun$; Fig.~\ref{fig:posterior_rout_mdot}),
the predicted NIR flux spans more than an order of magnitude across the mass range. JWST NIR filters (F070W--F444W) can therefore break the UV/optical mass degeneracy: low-mass accretors ($M \lesssim 5\,\Msun$) predict NIR emission well above the JWST sensitivity, while the high-mass models ($M \gtrsim 10\,\Msun$) are marginally detectable at best.

In the right panel of Fig.~\ref{fig:sed_comparison}, we show a zoom-in version of the prediction in the NIR-MIR bands. Other than the results from the fiducial He-dominated disk composition ($\kappa=0.2\rm\, cm^2\,g^{-1}$ and $\mu_{\rm e}=2$), we also show the predictions for solar composition with $\kappa=0.34\rm\, cm^2\,g^{-1}$ and $\mu_{\rm e}=1.2$. The differences between the two cases are significant (although within model uncertainties), and the physical origin of the differences and scatter will be explained in \S \ref{sec:ff_scatter}.

Our photometric predictions (median posterior) for the case of a He-dominated composition are listed in Table \ref{tab:jwst_predictions}.  Non-detection in the NIR bands would strongly disfavour low-mass accretors $M\lesssim 5M_\odot$; if the SED shows the predicted free-free excess, this would strongly support a wind origin\footnote{In the low accretion mass ($M \lesssim 5M_\odot$) wind emission scenario, another potentially testable prediction is that, as the accretion rate drops over a viscous timescale of the outer disk, we expect the wind color temperature to \textit{increase} over time (see eq. \ref{eq:Tw}) --- this is the opposite of the expectation from a viscously powered disk in which case we expect the effective temperature of the outer disk to \textit{decrease} as the disk expands over time \citep[see][]{2025arXiv251209017W}. } and constrain the accretor mass.


\begin{table*}
\centering
\caption{Predicted JWST AB magnitudes (median posterior) for the wind+irradiation model at $t = 1453\,\rm d$ and for a He-dominated composition with $\kappa = 0.2\,\rm cm^2\,g^{-1}$ and $\mu_{\rm e}=2$. For each filter, $\lambda_{\rm piv}$ is the pivot wavelength, and $m_{\rm lim}$ is the $5\sigma$ point-source limit for a $10\,\rm ks$ exposure. Predicted detections ($m_{\rm pred} < m_{\rm lim}$) are marked in boldface.}
\label{tab:jwst_predictions}
\begin{tabular}{llccccccc}
\hline\hline
Instr. & Filter & $\lambda_{\rm piv}$ ($\mu$m) & $m_{\rm lim}$ & $M=1.4\,M_\odot$ & $M=5\,M_\odot$ & $M=10\,M_\odot$ & $M=30\,M_\odot$ & $M=100\,M_\odot$ \\
\hline
NIRCam & F070W & 0.70 & 28.8 & $\mathbf{26.3}$ & $\mathbf{26.3}$ & $\mathbf{26.3}$ & $\mathbf{26.3}$ & $\mathbf{26.3}$ \\
NIRCam & F150W & 1.50 & 29.6 & $\mathbf{27.5}$ & $\mathbf{27.6}$ & $\mathbf{27.6}$ & $\mathbf{27.7}$ & $\mathbf{27.7}$ \\
NIRCam & F277W & 2.76 & 29.3 & $\mathbf{27.9}$ & $\mathbf{28.5}$ & $\mathbf{28.5}$ & $\mathbf{28.9}$ & $\mathbf{29.0}$ \\
NIRCam & F356W & 3.57 & 29.3 & $\mathbf{27.9}$ & $\mathbf{28.8}$ & $\mathbf{28.9}$ & $29.3$ & $29.5$ \\
NIRCam & F444W & 4.40 & 28.7 & $\mathbf{27.8}$ & $29.0$ & $29.1$ & $29.6$ & $29.9$ \\
\hline
MIRI & F560W & 5.60 & 26.9 & $27.8$ & $29.1$ & $29.3$ & $30.0$ & $30.3$ \\
MIRI & F770W & 7.70 & 26.2 & $27.8$ & $29.3$ & $29.5$ & $30.4$ & $30.9$ \\
\hline
\end{tabular}
\end{table*}

The full marginalized posterior distributions for all masses are shown in Appendix~\ref{app:full_posteriors}.

\subsection{Free-free emission: posterior scatter and composition dependence}
\label{sec:ff_scatter}

Below, we explain the scatter in the predicted NIR-MIR emission and its sensitivity to wind composition, as these directly affect potential JWST observations.

\textit{Large scatter at $M\lesssim5\,M_\odot$.}
For low-mass accretors the UV/optical is dominated by the wind photosphere
in the RJ regime.
In this limit the four HST bands constrain only the amplitude
$L_{\rm rad,tot}/T_{\rm w}^3\propto\kappa^{49/24}\dot{M}_{\rm d}^{55/24}f_{\rm col}^{-3}$,
leaving $\dot{M}_{\rm d}$ and $f_{\rm col}$ degenerate.
This degeneracy produces the broad $\dot{M}_{\rm d}$ posterior visible in
Fig.~\ref{fig:posterior_rout_mdot} at low masses and propagates directly
into a large scatter in the predicted free-free luminosity
$\nu L_{\rm ff, \nu}\propto\dot{M}_{\rm d}^2$ (eq.~\ref{eq:Lff}).

\textit{Tighter predictions at $M\gtrsim10\,M_\odot$.}
For higher-mass accretors ($10\lesssim M/\Msun\lesssim 100$) the UV/optical is dominated by disk irradiation, and the photospheric geometry provides an additional constraint as the reprocessing efficiency is of the order $r_{\rm ph}/\rd$. The observed SED nearly fixes $\rd$, and $r_{\rm ph}$ is also tightly constrained by the reprocessing geometry. Since $r_{\rm ph}\propto(\kappa\dot{M}_{\rm d})^{3/2}$ (eq.~\ref{eq:rph}), fixing $r_{\rm ph}$ pins $\kappa\dot{M}_{\rm d}=\text{const}$, breaking the $f_{\rm col}$--$\dot{M}_{\rm d}$ degeneracy and producing a narrower $\dot{M}_{\rm d}$ posterior and hence tighter free-free predictions (Fig.~\ref{fig:posterior_rout_mdot}).

\textit{Composition dependence.}
At fixed $\dot{M}_{\rm d}$, solar H-rich composition ($\kappa=0.34$, $\mu_{\rm e}=1.2$)
gives $\sim\!2\times$ more free-free luminosity (in the RJ limit) than the He-dominated
case ($\kappa=0.2$, $\mu_{\rm e}=2$) via $\nu L_{\rm ff,\nu}\propto\kappa^{2/3}\mu_{\rm e}^{-2/3}$ (eq.~\ref{eq:Lff}). For $M\lesssim5\,M_\odot$, this factor applies directly because $\dot{M}_{\rm d}$ is insensitive to $\kappa$ (the $f_{\rm col}$ degree of freedom absorbs the opacity change). For $M\gtrsim10\,M_\odot$, however, the geometric constraint forces
$\dot{M}_{\rm d}\propto\kappa^{-1}$, so $\nu L_{\rm ff,\nu}\propto\kappa^{2/3}\mu_{\rm e}^{-2/3}\times\kappa^{-2}=\kappa^{-4/3}\mu_{\rm e}^{-2/3}$;
evaluating this gives a factor of $\sim\! 0.7$, meaning solar composition
predicts $\sim\!30\%$ \textit{less} free-free at high masses.
The two effects therefore partially cancel, explaining why the right panel of
Fig.~\ref{fig:sed_comparison} shows similar NIR predictions for both
compositions at $M=10\,M_\odot$, whereas the low-mass ($M=1.4\,M_\odot$)
predictions differ by a factor of $\sim\!2$.

\subsection{(Lack of) Constraints on the accretor mass}
\label{sec:mass_constraints}

AT2018cow was detected by XMM-Newton in X-rays at $t\simeq 1350\rm\, d$ \citep{2024ApJ...963L..24M}. Due to potential contributions to the observed flux by other sources, we treat the detected flux as an upper limit\footnote{To be conservative, we take the $1\sigma$ band-integrated luminosity of $L_{0.3\mbox{-}10\mr{keV}} = 6\times10^{38}\rm\, erg\,s^{-1}$ \citep{2024ApJ...963L..24M}, whereas the spectral luminosity $\nu L_\nu$ near the lower end of the band pass is generally lower than $L_{0.3\mbox{-}10\mr{keV}}$. } $\nu L_\nu < 6\times10^{38}\rm\,erg\,s^{-1}$ at $\nu \simeq 10^{17}\rm\,Hz$ ($h\nu \simeq 0.4\rm\,keV$). This limit is plotted as a downward triangle in Fig.~\ref{fig:sed_comparison} and provides a useful constraint on the accretor mass. We note that \citet{2024ApJ...963L..24M} also provided hard X-ray upper limit $L_{10\mbox{-}79\rm\, keV} < 5\times10^{39}\rm\, erg\,s^{-1}$, which is less constraining for our model than the soft X-ray limit.

In the wind model, the photospheric color temperature
$T_{\rm w} = f_{\rm col}T_{\rm w}'$ scales as $T_{\rm w}'\propto \dot{M}_{\rm d}^{-17/24} M^{11/24}$ (eq.~\ref{eq:Tw}), so more massive accretors have hotter wind photospheres (the color correction factor $f_{\rm col}$ only makes $T_{\rm w}$ even higher).
For $M\gtrsim 100\,\Msun$ and the posterior accretion rates
($\dot{M}_{\rm d}\simeq2$--$3\times10^{-4}\,\msunyr$),
the wind color temperature reaches
$T_{\rm w}\sim (5$--$10)\times10^{5}\rm\,K$, shifting the blackbody peak
to soft X-ray frequencies and producing $\nu L_\nu\gtrsim 10^{39}\rm \,erg\,s^{-1}$
near $\nu\simeq10^{17}\rm\,Hz$ --- exceeding the XMM upper limit. We therefore disfavor $M\gtrsim 100\,\Msun$, with the caveat that the exact threshold depends on $f_{\rm col}$ and on whether the wind X-ray emission escapes quasi-isotropically through the low-mass ejecta inferred from the early light curve (see \S \ref{sec:ionization_breakout}).

Within the X-ray-allowed range $M \lesssim 100\,\Msun$, the UV/optical data
remain degenerate across all tested masses.
Wind free-free emission at NIR-MIR wavelengths
(eq.~\ref{eq:Lff}; Fig.~\ref{fig:sed_comparison})
provides a possible mass discriminant, as discussed in \S\ref{sec:SED_fits}.

\section{Discussion}

The wind+irradiation model relaxes the stringent mass constraints in the pure thin-disk picture and opens the accretor mass range down to $1.4\,\Msun$. We now discuss the connection to the micro-TDE formation scenario, a comparison with large-BH-mass interpretations, line emission from disk wind, the possible role of X-ray ionization breakout, and the main modeling uncertainties.

\subsection{Consistency with a micro-TDE origin}\label{sec:micro-TDEs}

We now show that the inferred disk parameters are potentially consistent with a
micro-tidal disruption event (micro-TDE) — the tidal disruption of a
main-sequence star by a compact object \citep{2016ApJ...823..113P}. Such a scenario may occur due to (1) a fortunate natal kick that leads to a close encounter between the newly formed compact object in a stripped-envelope SN and its companion star \citep{2025ApJ...986...84T}, or (2) in a merger between a compact object and its companion star's core in the case of a ``failed'' common envelope phase \citep{2025arXiv251009745K}. In this scenario, the debris disk is initially geometrically thick and
super-Eddington, and its accretion power is sufficient to power the LFBOT peak emission in the first days to months \citep[e.g.,][]{2019ApJ...881...75K}. The disk transitions to a geometrically thin state when its mass-inflow rate in the outer region drops below the Eddington rate, after which we expect the long-lived UV/optical plateau begins.


The transition to the thin state occurs when the accretion luminosity near the outer region of
the disk $r_{\rm d}$ drops below $L_{\rm Edd}$ \citep[e.g.,][]{2022hxga.book....3L}
\begin{equation}
    {GM\over r_{\rm d}}{M_{\rm d}\over t_{\rm vis,thick}} \simeq L_{\rm Edd}
    = {4\pi GMc\over\kappa},
    \label{eq:collapse_cond}
\end{equation}
where the viscous timescale in the thick state is
\begin{equation}
    t_{\rm vis,thick} = {r_{\rm d}^{3/2}\over\alpha h_{\rm thick}^2\sqrt{GM}},
    \label{eq:tvis_thick}
\end{equation}
where we have defined $$h_{\rm thick}\equiv H/r_{\rm d} \sim 0.5.$$
Substituting eq.~(\ref{eq:tvis_thick}) into (\ref{eq:collapse_cond}) gives
a constraint relating $r_{\rm d}$ and $M_{\rm d}$ at the thick-to-thin transition:
\begin{equation}
    r_{\rm d}^{5/2} = {\alpha h_{\rm thick}^2\kappa M_{\rm d}\sqrt{GM}\over 4\pi c}.
    \label{eq:rd_vs_Md}
\end{equation}

To obtain another constraint relating $r_{\rm d}$ and $M_{\rm d}$, we look at the current angular momentum of the disk
\begin{equation}
    J_{\rm d} \simeq \sqrt{GM\rd}\,M_{\rm d},
    \label{eq:Jd}
\end{equation}
which should be compared with the angular momentum budget from the \textit{initial} TDE for a star of mass $M_*$ and radius $R_*$
\begin{equation}
    J_{\rm TDE} \simeq \sqrt{GM r_{\rm t}}\,{M_*\over2},
    \quad r_{\rm t} = R_*\lrb{M\over M_*}^{1/3},
    \label{eq:JTDE}
\end{equation}
where we have assumed a circularization radius\footnote{As LFBOTs are more likely related to stellar-mass compact objects (rather than supermassive or intermediate-mass BHs), encounters with smaller pericenter radii $r_{\rm p} < r_{\rm t}$ are required for full disruptions and the corresponding stellar debris after the TDE have mildly eccentric orbits \citep{2022ApJ...933..203K}. These motivate the choice of a circularization radius of $r_{\rm t}$, instead of $\sim 2r_{\rm t}$ for TDEs by higher-mass BHs.} of $r_{\rm t}$ and that half the stellar mass stays bound.

In the super-Eddington thick-disk phase, disk winds likely carry mass and angular momentum away from the system, so the current disk angular momentum $J_{\rm d}$ (eq. \ref{eq:Jd}) is likely smaller than the initial value of $J_{\rm TDE}$ (eq. \ref{eq:JTDE}). We denote the ratio between the two by
\begin{equation}
    J_{\rm d} = f_{\rm j} J_{\rm TDE},
\end{equation}
where $f_{\rm j} < 1$ captures the uncertain angular momentum loss during the thick-disk phase. This leads to another relation between $r_{\rm d}$ and $M_{\rm d}$ at the thick-to-thin transition:
\begin{equation}\label{eq:Md_from_AM}
    M_{\rm d} = {1\over 2} f_{\rm j} \sqrt{r_{\rm t}/r_{\rm d}} M_*.
\end{equation}

Plugging eq. (\ref{eq:Md_from_AM}) into eq. (\ref{eq:rd_vs_Md}), we then obtain
\begin{equation}\label{eq:rd_cubic}
    r_{\rm d}^3 = {\alpha h_{\rm thick}^2\kappa f_{\rm j} \sqrt{GM r_{\rm t}} M_* \over 8\pi c},
\end{equation}
which gives
\begin{equation}\label{eq:rdthin}
    r_{\rm d,thin} \simeq 40.5R_\odot\, \lrb{\alpha h_{\rm thick}^2\over 10^{-2}}^{1/3} f_{\rm j,0.3}^{1/3} \kap^{1/3} \Mten^{2/9} \rstar^{1/6} \mstar^{5/18},
\end{equation}
where we have taken a fiducial value of $f_{\rm j}=0.3$ but it is rather uncertain.
This is the disk outer radius right after the thick-to-thin transition. The ratio between $r_{\rm d,thin}$ and the tidal disruption radius $r_{\rm t}$ is given by
\begin{equation}
    {r_{\rm d,thin}\over r_{\rm t}} \simeq 18.8\,\lrb{\alpha h_{\rm thick}^2\over10^{-2}}^{1/3}
    f_{\rm j,0.3}^{1/3} \kap^{1/3}\Mten^{-1/9}\rstar^{-5/6}
    \mstar^{11/18}.
\end{equation}
We find that, during the thick phase, the disk viscously spreads by about an order of magnitude; importantly, the ratio $r_{\rm d,thin}/r_{\rm t}$ depends weakly on all parameters --- especially for the tidal disruption of main-sequence stars with $r_*\propto m_*^{\approx 0.7}$.
The corresponding disk mass right after the thick-to-thin transition is given by
\begin{equation}
\begin{split}
    &M_{\rm d,thin} \simeq {f_{\rm j} \over 2}{\sqrt{r_{\rm t}/r_{\rm d,thin}} M_*}
    \simeq 3.5\times10^{-2}\Msun\, f_{\rm j,0.3}^{5/6} \\
    &\qquad \times \kap^{-1/6}  \Mten^{1/18} \rstar^{5/12} \mstar^{25/36} \lrb{\alpha h_{\rm thick}^2\over10^{-2}}^{-1/6}.
\end{split}
    \label{eq:Mdthin}
\end{equation}
The collapse occurs on the viscous timescale evaluated at $r_{\rm d,thin}$ during the \textit{thick phase}:
\begin{equation}
\begin{split}
    &t_{\rm thin} \sim t_{\rm vis,thick}(r_{\rm d,thin})
    = {r_{\rm d,thin}^{3/2}\over\alpha h_{\rm thick}^2\sqrt{GM}}\\
    &\simeq 0.4\,\mr{yr}\,f_{\rm j,0.3}^{1/2} \kap^{1/2}\Mten^{-1/6}\mstar^{5/12}
    \rstar^{1/4} \lrb{\alpha h_{\rm thick}^2\over10^{-2}}^{-{1\over 2}}.
\end{split}
    \label{eq:tthin}
\end{equation}

We caution that the hydrodynamics of micro-TDEs are more complicated than the simplified model described here. For instance, the initial disk mass strongly depends on the pericenter separation of the encounter, and in most cases, stars undergo multiple partial disruptions with decreasing orbital periods \citep[e.g.,][]{2022ApJ...933..203K, 2024A&A...685A..45V}. In the merger scenario of \citet{2025arXiv251009745K}, the initial disk mass depends on the mass ejection prior to and during the dynamically unstable phase, and the initial disk radius also depends on the angular momentum transport during the merger. We do not consider these details in this work, as our goal is to demonstrate the broad consistency of the AT2018cow data with the micro-TDE picture. Future work should explore the sensitivity of the results on the micro-TDE hydrodynamics.

A detailed disk evolution calculation, including the angular momentum loss during this process (parameterized by $f_{\rm j}$), is beyond the scope of this work, but we note that the evolutionary time for the thick disk to viscously spread to $r_{\rm d,thin}$ is expected to be somewhat shorter than $t_{\rm vis,thick}(r_{\rm d,thin})$. Despite the uncertainties, we find that the collapse time of a micro-TDE disk is consistent with the plateau emerging within $\sim1\,\rm yr$.

After the transition, the disk aspect ratio drops from $h_{\rm thick}\sim0.5$
to $h_{\rm thin}\ll h_{\rm thick}$, so the thin-state viscous timescale is
\begin{equation}
    t_{\rm vis,thin}
    \sim 10^2\,t_{\rm thin}\lrb{h_{\rm thick}\over 10\,h_{\rm thin}}^2,
    \label{eq:tvis_thin}
\end{equation}
yielding a long-lived plateau lasting for $\sim10\,\rm yrs$ or longer.
The accretion rate in the thin-disk state is then given by
\begin{equation}
\begin{split}
    &\dot{M}_{\rm d,thin} = {M_{\rm d,thin}\over t_{\rm vis,thin}}\simeq 3.5\times10^{-3}\msunyr\,f_{\rm j,0.3}^{5/6}  \\
    &\quad \times  \kap^{-1/6} \Mten^{1/18} \rstar^{5/12} \mstar^{25/36} \lrb{t_{\rm vis,thin}\over10\,\mr{yr}}^{-1} \lrb{\alpha h_{\rm thick}^2\over10^{-2}}^{-{1\over 6}}.
\end{split}
    \label{eq:Mdotthin}
\end{equation}

Fig.~\ref{fig:posterior_rout_mdot} overlays the micro-TDE predictions (orange band) on the posterior distributions of $r_{\rm d}$ and $\dot{M}_{\rm d}$, evaluated from equations~(\ref{eq:rdthin}) and (\ref{eq:Mdotthin}) for a solar-type disrupted star ($M_*=M_\odot$ and $R_*=R_\odot$), $\alpha h_{\rm thick}^2\in[10^{-3},\,0.1]$, $t_{\rm vis,thin}\in[5,\,30]\,\rm yr$, and $f_{\rm j}\in [0.1, 1]$. The predicted $r_{\rm d,thin}$ is broadly consistent with the posterior disk radii; the predicted $\dot{M}_{\rm d, thin}$ slightly exceeds the posterior values for the highest compact object masses $M\gtrsim 10^2M_\odot$ (which are disfavored by the X-ray constraints anyway). We conclude that the disk powering the UV/optical plateau in AT2018cow is consistent with that expected from a micro-TDE.

\subsection{Comparison with large-BH-mass interpretations}
\label{sec:comparison}

Recent analyses of the AT2018cow late-time UV/optical plateau by
\citet{2025MNRAS.544L.108I} and \citet{2025arXiv251209017W}
attribute the emission purely to viscous heating\footnote{We note that \citet{2025arXiv251209017W} cautiously mentioned the possibility of irradiation-powered emission in the UV/optical without a detailed calculation.} in a geometrically thin disk; we showed in \S\ref{sec:thin_disk} that disk
self-consistency requirements (Fig.~\ref{fig:Md_rd_constraints_main}) constrain $M\gtrsim 200\,\Msun\,\alpha_{-2}^{2/3}\kap^{4/3}$. This is in tension with the XMM-Newton X-ray constraint, unless one considers unusually low viscosity $\alpha\ll 0.01$. Our wind+irradiation model sidesteps the minimum-mass requirement entirely: the pressure constraint (eq.~\ref{eq:pressure_constrant_csr2_csg2_cs2}) does not apply once the viscously powered emission becomes subdominant.

For low-mass accretors $1.4\lesssim M/\Msun \lesssim 5$, the UV/optical emission is dominated by the wind emission and the wind photospheric color depends only weakly on $M$ (eq.~\ref{eq:Tw}). For $10\lesssim M/\Msun \lesssim 100$, the UV/optical emission is dominated by reprocessing of the wind emission by the outer disk --- in this case, the outer disk radius is well constrained by the observed SED (eq. \ref{eq:R_BB}) but, as the effective temperature $T_{\rm eff}(\rd)$ is largely set by irradiation, the midplane temperature from viscous heating alone is much less than $\tau^{1/4}T_{\rm eff}(\rd)$ (eq. \ref{eq:Tmid}), relaxing the radiation pressure constraint. Only for $M\gtrsim 300\Msun$, our model posterior settles on a pure viscously powered thin disk solution.

Our analysis shows that, for lower-mass accretors, the wind emission and irradiation are physically important and cannot be ignored in a self-consistent model. We conclude that the UV/optical data cannot distinguish between a neutron star and a BH with, e.g., $M=10\,\Msun$.

\subsection{Line emission from H- or He-rich disk winds}

A potential way to differentiate between H- and He-rich wind compositions is by the emission lines from the disk wind. In the following, we estimate the recombination line luminosities assuming fully ionized wind. For a H-dominated composition, we focus on the H$\alpha$ line; and for the He-dominated case, we focus on the HeII1640 line (the other promising line is HeII4686, but its luminosity is much smaller due to lower line energy and branching ratio).

In the optically thin limit\footnote{Even if the line center is moderately optically thick due to bound-bound absorption (if the $n=2$ quantum states are sufficiently populated), the line photons will simply resonantly scatter their way out in the Case-B limit and hence the line luminosity is given by the optically thin limit.}, the emission rate of H$\alpha$ or HeII1640 from Case-B recombination cascade above the continuum photospheric radius $r_{\rm ph}$ can be roughly estimated by
\begin{equation}
    {L_{\rm line}\over \epsilon_{\rm line}} \simeq \int_{r_{\rm ph}}^\infty 4\pi r^2 \alpha_{\rm B} f_{\rm br} n_{\rm e} n_{\rm i} \d r
    \simeq \alpha_{\rm B} f_{\rm br} {X_{\rm i}\dot{M}_{\rm d}/v\over \kappa \mu_{\rm e} A \mproton^2},
\end{equation}
where we have assumed H and He to be fully ionized (into HII and HeII) with electron and ion number densities $n_{\rm e} = \rho/(\mu_{\rm e}\mproton)$ and $n_{\rm i} = X_{\rm i} \rho/(A \mproton)$ ($X_{\rm i}$ being the ion mass fraction and $A$ being the ion mass number), $f_{\rm br}\simeq 1/3$ is the branching ratio for H$\alpha$ and HeII1640 lines (ignoring temperature dependence), and $\alpha_{\rm B}$ is the Case-B recombination rate coefficient \citep{2011piim.book.....D}
$$
\alpha_{\rm B} \simeq 1.4\times10^{-13} \mr{\,cm^3\,s^{-1}}\, Z\lrb{T_{\rm e,4.3}/Z^2}^{-0.8},
$$
$T_{\rm e} = 10^{4.3}T_{\rm e,4.3}\rm\, K$ is the electron temperature, and $Z$ is the nuclear charge. Putting in the wind velocity $v(r_{\rm sph}) \simeq \sqrt{GM/r_{\rm sph}} = \sqrt{L_{\rm Edd}/\dot{M}_{\rm d}}$ and line photon energy $\epsilon_{\rm line} = \lrb{5/36}Z^2\mr{Ry}$, we find
\begin{equation}\label{eq:line_luminosity}
\begin{split}
    &L_{\rm line} \simeq 2.5\times10^{38}\mr{\,erg\,s^{-1}}\,{X_{\rm i}(Z/2)^{2.6}\over \kap^{1/2}(\mu_{\rm e}/2)} {\dot{M}_{\rm d,-3}^{3/2}\over T_{\rm e,4.3}^{0.8}  \Mten^{1/2}}\\
    &\quad \simeq 3.7\times10^{37}\mr{\,erg\,s^{-1}}\,{(X_{\rm i}/0.7)Z^{2.6}\over \kappa_{0.34}^{1/2} (\mu_{\rm e}/1.2)} {\dot{M}_{\rm d,-3}^{3/2}\over T_{\rm e,4.3}^{0.8}  \Mten^{1/2}}.
\end{split}
\end{equation}
where $\dot{M}_{\rm d,-3} = \dot{M}_{\rm d}/10^{-3}\,\msunyr$ and the fiducial values are for a He-dominated wind composition. For H-dominated solar composition ($X_{\rm i}=0.7$, $Z=1$, $\mu_{\rm e}=1.2, \kappa=0.34\mr{\,cm^2\,g^{-1}}$), we get a pre-factor of $X_{\rm i}(Z/2)^{2.6}(\mu_{\rm e}/2)^{-1}\kap^{-1/2}\approx 0.15$ --- for the same $\dot{M}_{\rm d}$ and $T_{\rm e}$, the H$\alpha$ luminosity for the H-dominated case is lower than the HeII1640 luminosity for the He-dominated case (but $T_{\rm e}$ is likely higher for the He-dominated case). The line width is given by the outflow velocity
\begin{equation}\label{eq:wind_velocity}
    v(r_{\rm sph}) \simeq 2.0\times10^3\mr{\, km\,s^{-1}}\, \lrb{\Mten/\kap \over \dot{M}_{\rm d}/10^{-3}\,\msunyr}^{1/2}.
\end{equation}

\begin{table*}
\centering
\caption{Predicted recombination line luminosities $\log_{10}(L_{\rm line}/\rm erg\,s^{-1})$ (median and $1\sigma$ range) for the wind+irradiation model at $t = 1453\,\rm d$, assuming $T_{\rm e}=10^{4.3}\,\rm K$. HeII1640 assumes a He-dominated wind ($\kappa=0.2\,\rm cm^2\,g^{-1}$, $\mu_{\rm e}=2$, $X_{\rm i}=1$); H$\alpha$ assumes a H-dominated solar-composition wind ($\kappa=0.34\,\rm cm^2\,g^{-1}$, $\mu_{\rm e}=1.2$, $X_{\rm i}=0.7$), each evaluated using its corresponding $\dot{M}_{\rm d}$ posterior (eq.~\ref{eq:line_luminosity}).}
\label{tab:line_predictions}
\renewcommand{\arraystretch}{1.4}
\begin{tabular}{lcccccc}
\hline\hline
Line & $\lambda$ (\AA) & $M=1.4\,M_\odot$ & $M=5\,M_\odot$ & $M=10\,M_\odot$ & $M=30\,M_\odot$ & $M=100\,M_\odot$ \\
\hline
H$\alpha$ & 6563 (air) & $38.02^{+0.41}_{-0.52}$ & $37.45^{+0.80}_{-0.47}$ & $36.94^{+0.35}_{-0.25}$ & $36.68^{+0.36}_{-0.24}$ & $36.09^{+0.20}_{-0.11}$ \\
HeII1640 & 1640 & $38.86^{+0.61}_{-0.50}$ & $38.16^{+0.43}_{-0.27}$ & $37.96^{+0.27}_{-0.23}$ & $37.53^{+0.51}_{-0.17}$ & $37.04^{+0.14}_{-0.10}$ \\
\hline
\end{tabular}
\end{table*}

The predicted median luminosities of HeII1640 and H$\alpha$ line emission from the disk wind for He- and H-dominated composition (respectively) are shown in Table \ref{tab:line_predictions}. We find that the predicted line luminosities decrease rapidly with mass of the accretor and could be used to constrain the accretor mass and disk composition.

Late-time constraints of H$\alpha$ line emission have been obtained based on narrowband imaging with the HST F665N filter at rest-frame time $t\simeq 703\rm\, d$ \citep{2022MNRAS.512L..66S,2023ApJ...955...43C}. At $z=0.014$, the H$\alpha$ rest wavelength in vacuum shifts to 6656$\rm \AA$, precisely within the F665N filter (whereas the other filter F657N misses the redshifted H$\alpha$ line and gives an unsurprising non-detection). The wind velocity of $v\sim 2000\rm\, km/s$ (eq. \ref{eq:wind_velocity}) corresponds to a full-width-half-maximum of $\sim 100\rm\,\AA$ (for a Gaussian with $\sigma=v$), which matches with the effective rectangular width of the F665N filter (131\AA). Although the much narrower ($\sigma\sim 20\rm\, km\,s^{-1}$) H$\alpha$ and [NII] lines from the nearby HII region \citep{2023MNRAS.519.3785S} also contribute to the F665N flux, the high spatial resolution of HST allows spatial separation of the contribution from AT2018cow (despite the lack of wavelength resolution). This makes it possible to use the F665N image to directly measure/constrain the H$\alpha$ luminosity, which in turn constrains the accretor mass and disk composition.

Unfortunately, the HST F665N image is somewhat too shallow for a definite detection. \citet{2022MNRAS.512L..66S} reported a marginal detection at $4.6\sigma$ whereas \citet{2023ApJ...955...43C} only found excess emission at the level of $2.8\sigma$ using the same data. If we take the measurement by \citet{2022MNRAS.512L..66S} (24.47$\pm$0.24 Vega mag), the inferred in-band flux is $4\pm 1 \times10^{-17}\rm\, erg\,s^{-1}\,cm^{-2}$, which corresponds to an in-band luminosity of $2\pm 0.5\times 10^{37}\rm\, erg\,s^{-1}$. However, this should not be taken as the H$\alpha$ line flux, because (1) part of the line may be outside the F665N band (which is the case for a Gaussian-broadened line with $\sigma\gtrsim 1500\rm\, km\,s^{-1}$), and (2) contribution from continuum emission needs to be subtracted (which depends on the continuum model). It is beyond the scope to carefully constrain the H$\alpha$ line luminosity. If the H$\alpha$ detection is real, this supports the picture of the tidal disruption of a main-sequence companion star, instead of a He star. If we assume a line luminosity of $L_{\rm line}\sim 1\times10^{37}\rm\, erg\,s^{-1}$, this would be consistent with a BH accretor with $5M_\odot \lesssim M \lesssim 30M_\odot$ and in tension with a neutron star accretor ($M=1.4M_\odot$) or a very massive BH accretor ($M\gtrsim 30M_\odot$) at $1\sigma$ confidence level.

We note that \citet{2023MNRAS.519.3785S} presented VLT/MUSE-IFU observation of AT2018cow covering the wavelength range of 4750-9350$\AA$ (resolution of 2000-3000) at $\sim\!1\rm\, yr$ post discovery. Their observing mode had angular resolution (PSF) of $0.86''$, which may be insufficient to isolate the small contribution of disk wind H$\alpha$ flux from the nearby star-forming complex --- they measured the H$\alpha$ luminosity from the nearby HII region (``Region 0'') to be $3.4\times10^{38}\rm\, erg\,s^{-1}$. However, the disk wind H$\alpha$ line is much broader than that from the HII region, so it is in principle possible to constrain/measure the disk wind H$\alpha$ emission using archival MUSE data.

Finally, we also note that, at earlier time $t\sim 30\,$d, the observed H$\alpha$ line luminosity of $L_{\rm H\alpha}\gtrsim 10^{39}\rm\, erg\,s^{-1}$ \citep{2019ApJ...872...18M} may be consistent with the emission from the disk wind provided that the accretion ($\approx$outflow) rate was much higher, $\dot{M}_{\rm d}\gtrsim 10^{-2}\,\msunyr$. If the H$\alpha$ line originates from the disk wind, it supports the scenario of a H-rich composition. It is possible that the disk radius $r_{\rm d}$ is less than $r_{\rm sph}$ and, in this case, the wind velocity is correspondingly higher than used in eq. (\ref{eq:line_luminosity}) by a factor of $\sqrt{r_{\rm sph}/r_{\rm d}}$, which slightly reduces the predicted line luminosity by this factor.

\subsection{Soft X-ray ionization breakout from the ejecta}\label{sec:ionization_breakout}

Soft X-rays from the disk or wind may be absorbed by bound K-shell electrons of Carbon and Oxygen (and L-shell electrons of Fe) inside the ejecta launched near the peak luminosity. In the following, we consider bound-free absorption of X-rays by an ejecta of mass $M_{\rm ej}$, assuming a significant fraction of C/O composition\footnote{If the ejecta is dominated by H or He with low metal abundance, the soft-X-ray bound-free opacity is much smaller than for C/O-rich ejecta.}. Near the time of XMM-Newton observation, the ejecta is located at a characteristic radius of $r_{\rm ej} = v t\sim 5\times10^{17}\mr{\,cm}$ for $v\sim 0.1c$ and $t\sim 5\rm\, yr$. Let us consider a simplified model where the ejecta is spherically symmetric and has a uniform density $\rho_{\rm ej} = 3M_{\rm ej}/(4\pi r_{\rm ej}^3)$. 

For a given luminosity of ionizing photons $L_{\rm ion} =\int_{\epsilon_{\rm ion}}^\infty L_\nu \d \nu$ (where $\epsilon_{\rm ion}=$ ionization threshold for a certain species, e.g., Carbon) and a given ejecta radius $r_{\rm ej}$, there is a critical ejecta mass $M_{\rm ej,crit}$ at which the ejecta is marginally fully ionized. If the ejecta mass is greater than $M_{\rm ej, crit}$, we are in the ionization-bounded limit: there will be a neutral layer in the outer regions of the ejecta and, in this case, the escaping ionizing X-ray luminosity is much below the incident one --- the XMM-Newton X-ray upper limit does not strongly constrain the accretor mass in this case. On the other hand, if the ejecta mass is smaller than $M_{\rm ej, crit}$, we are in the density-bounded limit: the ejecta is fully ionized and the escaping ionizing X-ray luminosity is very close to the incident one.

To find the critical ejecta mass $M_{\rm ej,crit}$ for ionization breakout, we consider that the ionization rate is balanced with recombination rate in a fully ionized ejecta
\begin{equation}
    {L_{\rm ion}\over \lara{\Delta \epsilon}} = N_{\rm CO} \alpha_{\rm B} n_{\rm e},
\end{equation}
where $\lara{\Delta \epsilon}$ is the average cost per ionization (hereafter taken to be $\sim\!1\rm\, keV$), $N_{\rm CO} = f_{\rm CO} M/(A\mproton)$ is the total number of C/O nuclei relevant for recombination ($A$ being the average mass number, $f_{\rm CO}$ is C$+$O mass fraction), $n_{\rm e} = \rho_{\rm ej}/(\mu_{\rm e}\mproton)$ is the electron number density, $\mu_{\rm e}$ is the mean molecular weight per electron.
For X-ray ionizations, the corresponding recombinations are dominated by H-like C/O, so we use $\mu_{\rm e} \approx 2$, $A\approx 2Z$, $T_{\rm e}\simeq 10^5\rm\, K$, $6\lesssim Z\lesssim 8$, $\lara{\Delta \epsilon}\sim 1\rm\, keV$, and hence $(T_{\rm e,4.3}/Z^2)^{0.4}\simeq 0.4$. Thus, we find the critical ejecta mass
\begin{equation}\label{eq:Mejcrit}
    \begin{split}
        &M_{\rm ej,crit} = \lrb{{L_{\rm ion}\over \lara{\Delta \epsilon}}{4\pi r^3 \mu_{\rm e} A m_{\rm p}^2 \over 3f_{\rm CO}\alpha_{\rm B} }}^{1/2}\\
&\simeq 1.9 M_\odot\, r_{17.7}^{3/2} L_{\rm ion,39}^{1/2} \lara{\Delta \epsilon}_{\rm keV}^{-1/2} \lrb{f_{\rm CO}\over 0.5}^{-1/2}  .
    \end{split}
\end{equation}
Fig. \ref{fig:sed_comparison} shows that, for accretor masses $M\gtrsim 100M_\odot$, the ionizing luminosity above the C-ionization threshold of $\simeq0.5\rm\, keV$ (or 25\AA) is in the range of $L_{\rm ion}\sim 10^{39}$--$10^{40}\rm\, erg\,s^{-1}$. The soft X-rays will only be substantially absorbed if $M_{\rm ej}\gtrsim 2$ to $6M_\odot$.

The early-time UV/optical lightcurve of AT2018cow indicates a diffusion timescale of a few days and hence a relatively low ejecta mass $M_{\rm ej}\lesssim 0.5M_\odot$ \citep[e.g.,][]{2019ApJ...872...18M}. In this work, we assume that the ejecta mass is indeed this low \citep[as favored by][]{2026arXiv260118887G} and hence the soft X-rays largely escape from the ejecta as long as $L_{\rm ion}\gtrsim 10^{38}\rm\, erg\,s^{-1}$. Therefore, the XMM-Newton flux limit can be used to rule out accretor masses $M\gtrsim 100M_\odot$.

However, the caveat is that the system may be highly non-spherical such that a large fraction of solid angle is covered by ``dark mass'' with lower velocities and located at much smaller radii $\ll 5\times10^{17}\rm\,cm$ \citep[as is the case in the model by][]{2025arXiv251009745K}. In that case, it is possible that the X-ray emission from the disk wind has been substantially absorbed, and hence the XMM-Newton limit does not strongly constrain the accretor mass (meaning that $M\gtrsim10^2M_\odot$ is allowed). A potential discriminator in this case is the spectrum of the reprocessed emission by the absorber, which, however, is beyond the scope of this work.

\subsection{Uncertainties}

Our model makes several simplifying assumptions, which introduce uncertainties in the predicted SED. We discuss these uncertainties in this section.

\textit{Wind spectrum.}
We model the wind emission as a blackbody at the color temperature $T_{\rm w} = f_{\rm col}T_{\rm w}'$, with $f_{\rm col}$ as a free parameter. In reality the spectrum may depart significantly from a Planck function: a scattering-dominated photosphere produces a hardened continuum with $f_{\rm col} > 1$, and recombination can add emission lines (most prominently H$\alpha$ if H-rich) that are not captured by a single-temperature blackbody. Detailed predictions would require full radiative-transfer calculations through the wind \citep[e.g.,][]{2026ApJ..1001..191A}, which is beyond the scope of this work. The color correction $f_{\rm col}$ absorbs most of this uncertainty
at UV/optical wavelengths, but our model cannot predict the line emission.

\textit{Outer-disk irradiation geometry.}
Three geometric idealizations affect the irradiation calculation.
(i) We set the disk albedo to zero; a realistic non-zero albedo
(reflecting a fraction of the incident wind radiation) would reduce the
irradiation-powered disk luminosity by a factor of $1-\varpi$, where $\varpi$ is the
disk albedo.
(ii) We treat the disk as a geometrically flat sheet; in reality the
disk surface is likely flared ($H/r$ increasing outward), which would
moderately enhance irradiation at large radii.
(iii) We assume an infinitely sharp wind photosphere; in practice the
electron-scattering optical depth decreases smoothly with radius, so
photons scattered in the extended wind contribute additional irradiation
of the outer disk. Each of these effects modifies the disk luminosity by a factor of
order unity, with partial compensation.

\textit{Compact star cluster contribution to NIR-MIR.} The possibility that AT2018cow is produced by a TDE by an intermediate-mass BH is allowed by our analysis, provided that we relax the XMM-Newton constraint by considering absorption of soft X-rays (see \S \ref{sec:ionization_breakout}). In this scenario, we expect the intermediate-mass BH to be located inside a dense star cluster, which will enhance the TDE rate by orders of magnitude as compared to the case of a free-floating BH. In fact, an underlying young star cluster of mass up to $10^3$--$10^4M_\odot$ (age dependent) is allowed by HST photometry, as has been discussed by \citet{2023MNRAS.525.4042I,2025MNRAS.544L.108I}. This adds further complications to the wind free-free emission interpretation, should JWST detect the flux excess in the NIR-MIR above the RJ extrapolation. The best way of breaking the degeneracy between disk+wind and star cluster contribution would be to study the long-term evolution of AT2018cow. As the accretion rate drops, we expect the wind free-free emission to fade over the viscous timescale of the outer disk ($\sim$10 yrs).

\textit{SN ejecta contribution to NIR-MIR.}
We have ignored emission and absorption from the SN ejecta at radii well beyond $r_{\rm ph}$.
At $t\gtrsim 1\rm\,yr$ the ejecta is largely transparent at UV/optical
wavelengths, but the ionized region of the ejecta may contribute free-free emission in the NIR-MIR, which would complicate the interpretation of JWST photometry as \textit{disk wind} free-free emission. Below, we estimate this contribution. To be conservative, we assume that the ejecta is in the ionization-bounded regime (such that the ionizing photons are largely absorbed). If, instead, the ejecta is in the density-bounded regime (as is the case for a low ejecta mass $M_{\rm ej}\lesssim 0.5\Msun$), the free-free luminosity would be much lower than what is estimated below.

In the ionization-bounded regime, the mass of the ionized region is given by
\begin{equation}
    M_{\rm ion} = {L_{\rm ion}\over \lara{\Delta \epsilon}} {A\mproton \over \alpha_{\rm B} n_{\rm e}}.\ \mbox{ (if ionization-bounded)}
\end{equation}
The free-free \textit{mass emissivity} is given by (ignoring the Gaunt factor)
\begin{equation}
    j_{\rm ff,\nu}/\rho_{\rm ej} = \kappa_{\rm ff,\nu} B_\nu = {2h Z^2 Q n_{\rm e}\over A c^2T_{\rm e}^{1/2}} \mr{e}^{-h\nu/\kB T_{\rm e}}.
\end{equation}
Hereafter, we take $\mr{e}^{-h\nu/\kB T_{\rm e}}\approx 1$ as we are interested in the NIR-MIR bands for which $h\nu\ll \kB T_{\rm e}$. Thus, in the ionization-bounded regime, the free-free spectral luminosity is given by
\begin{equation}\label{eq:Lff_ejecta}
\begin{split}
    &\nu L_{\rm ff,\nu} = {\nu j_{\rm ff,\nu}\over \rho_{\rm ej}} M_{\rm ion} \approx {L_{\rm ion}\over \lara{\Delta \epsilon}} {2h\nu Q Z^2\mproton \over c^2 \alpha_{\rm B}T_{\rm e}^{1/2}}\\
    &\approx 1.6\times10^{35}\mr{\,erg\,s^{-1}}\, \nu_{14} {L_{\rm ion,39}\over \lara{\Delta \epsilon}_{\rm 100eV}}  (T_{\rm e,4.3}/Z^2)^{0.3},
\end{split}
\end{equation}
where the $(T_{\rm e,4.3}/Z^2)^{0.3}\sim 1$ depends weakly on the ejecta composition and we have taken a fiducial average ionization cost of $\lara{\Delta \epsilon} \sim 100\rm\, eV$ for a He-dominated ejecta composition (and He nuclei dominating the free-free emission). For C/O-dominated ejecta composition, we expect $\lara{\Delta \epsilon} \sim 1\rm\, keV$, and the corresponding free-free emission is much weaker. For a H-dominated ejecta composition, we expect the free-free luminosity to be higher by a factor of a few due to lower average ionization cost $\lara{\Delta \epsilon}$, but the system may be in the density-bounded regime as $M_{\rm ej,crit}$ (eq. \ref{eq:Mejcrit}) is also larger.

Comparing the free-free emission from the ejecta (eq. \ref{eq:Lff_ejecta}) with that from the disk wind (eq. \ref{eq:Lff}), we find that the ejecta contribution should be minor in the low accretor mass cases. For the $M=1.4M_\odot$ case (neutron star accretor), the disk wind free-free emission has $\nu L_{\rm ff,\nu}\sim 10^{37}\rm\, erg\,s^{-1}$ near $\nu=10^{14}\rm\, Hz$ (or $\lambda\simeq 3\rm\,\mu m$); for $M=5M_\odot$, we expect $\nu L_{\rm ff,\nu}\sim 10^{36}\rm\, erg\,s^{-1}$ from the disk wind. However, at $M\gtrsim10M_\odot$, the disk wind free-free emission becomes much weaker and is comparable to the ejecta free-free emission if the latter is in the ionization-bounded limit. Thus, we find that it is difficult to distinguish between these two contributions for $M\gtrsim 10M_\odot$; the free-free excess is difficult to detect with JWST anyway. If the free-free excess is indeed detected, since the ionized ejecta is optically thin to its own free-free emission, we expect a flatter NIR-MIR spectrum $L_\nu\propto \nu^0$ from ejecta free-free emission rather than the $L_\nu\propto \nu^{2/3}$ power-law expected from the disk wind.

\section{Summary}\label{sec:summary}

We have proposed a super-Eddington wind emission plus outer-disk irradiation model
for the late-time UV/optical plateau of AT2018cow and fitted it to
the four-band HST photometry at $t=1453\,\rm d$.
The main results are as follows.

\begin{enumerate}

\item \textit{Wind+irradiation model.}
When the disk accretion rate exceeds the global Eddington value, the inner disk
($r<r_{\rm sph}$) launches an optically thick wind whose photosphere emits
a blackbody spectrum at color temperature
$T_{\rm w}=f_{\rm col}T_{\rm w}'\propto\dot{M}_{\rm d}^{-17/24}M^{11/24}$
(eq.~\ref{eq:Tw}). The geometrically thin outer disk region beyond the wind photosphere ($r_{\rm ph}<r<\rd$) is simultaneously heated by viscous accretion and irradiated by the wind photosphere, contributing additional UV/optical flux (§\ref{sec:wind_model}). At NIR-MIR wavelengths the ionized wind produces free-free emission
with $\nu L_\nu\propto\nu^{5/3}\dot{M}_{\rm d}^2 M^{-2/3}$ (eq.~\ref{eq:Lff}).

\item \textit{UV/optical mass degeneracy.}
All compact object masses from $1.4\,\Msun$ (NS) to
$\sim100\,\Msun$ produce qualitatively good fits to the four-band HST data,
with log-evidences differing by at most $\Delta\ln\mathcal{Z}\simeq2.6$
— inconclusive on the Jeffreys scale (\S \ref{sec:SED_fits}).
The reason is that the wind photospheric color depends only weakly on $M$,
and the color correction factor $f_{\rm col}\in[1.6,3.7]$ absorbs the
remaining variation. This relaxes the requirement of a high-mass BH
$M\gtrsim 200\Msun\,\alpha_{-2}^{2/3} \kap^{4/3}$ in the viscously powered thin-disk model \citep{2025MNRAS.544L.108I, 2025arXiv251209017W}.

\item \textit{Constraints on the accretor mass.}
Under the assumption that the soft X-ray wind emission escapes quasi-isotropically, the XMM-Newton limit disfavors $M\gtrsim100\,\Msun$: at such masses the wind photosphere is hot  enough ($T_{\rm w}\gtrsim5\times10^5\,\rm K$) to radiate significantly in the soft X-ray band (§\ref{sec:mass_constraints}). This leaves a broad range of accretor masses allowed by HST observations $1.4\lesssim M/\Msun\lesssim100$.
JWST NIR-MIR photometry can break the UV/optical mass degeneracy: the free-free luminosity at $\lambda\sim1$--$10\,\mu\rm m$ differs by an order of magnitude between neutron star and $\gtrsim 10\,\Msun$
black hole models (Fig.~\ref{fig:sed_comparison}). Another potential way to break the mass degeneracy is to constrain the HeII1640 and H$\alpha$ line luminosities via UV and optical spectroscopy (respectively): the predicted line luminosities decrease rapidly with the accretor mass (Table \ref{tab:line_predictions}).

\item \textit{Consistency with micro-TDEs.}
The inferred disk parameters are broadly consistent with a thin disk that forms following a micro-TDE — the tidal disruption of a star by a compact object either due to a nearly parabolic encounter or a merger following dynamically unstable mass transfer (§\ref{sec:micro-TDEs}; Fig.~\ref{fig:posterior_rout_mdot}). The initially thick disk transitions to a thin one on a timescale of $t_{\rm thin}\simeq0.1$--$1\,\rm yr$ (eq.~\ref{eq:tthin}), consistent with the UV/optical plateau rising within $\sim1\,\rm yr$ of the explosion, after which the thin-state disk evolves on the much longer
viscous timescale $\gtrsim\!10\,\rm yr$ that sustains the long-lived plateau.

\end{enumerate}

Taken together, our results suggest that AT2018cow is most naturally understood as an accretion-powered transient in which a compact object --- either a neutron star or a stellar-mass black hole --- undergoes a dynamical interaction with a companion star or its helium core, producing a long-lived accretion disk. The initially super-Eddington accretion powers the luminous fast blue optical transient peak via an energetic wind, and the subsequent sub-Eddington thin outer disk, irradiated by its own wind emission, sustains the slowly-fading UV/optical plateau for $\sim\!10\,\rm yrs$ or longer. This picture resolves the tension between the large accretor mass required by pure thin-disk models and the upper limit implied by the XMM-Newton soft X-ray constraint. Two specific formation channels are consistent with the inferred disk parameters: a nearly parabolic tidal disruption of a main-sequence companion \citep{2025ApJ...986...84T}, and a merger with a companion's helium core during dynamically unstable mass transfer \citep{2025arXiv251009745K}. These scenarios produce disks with different compositions (affecting $\kappa$ and $\mu_{\rm e}$), but our qualitative conclusions are unchanged.

The central open question --- the identity of the compact object --- cannot be answered by UV/optical photometry alone. A potentially decisive test is to use JWST to detect or rule out the wind NIR-MIR free-free emission from a neutron star accretor (testable with a modest $10\,\rm ks$ exposure, Table~\ref{tab:jwst_predictions}); for a black hole accretor with $M\gtrsim 10M_\odot$, the disk wind would produce much weaker free-free emission. The composition dependence of the free-free luminosity (a factor of $\sim\!2$ between He- and H-dominated winds for a neutron star accretor; \S\ref{sec:ff_scatter}) may further discriminate the disk composition and thereby shed light on the formation channel. Whether AT2018cow harbors a neutron star or a stellar-mass black hole, it represents a new class of decade-long accretion transients accessible only through late-time UV monitoring, and a compelling case for JWST follow-up.

\section*{Acknowledgment}

We thank Anna Ho, Raffaella Margutti, Eliot Quataert, Yuhan Yao, Savannah Cary, and Stephon Qian for insightful comments and suggestions. W. L.'s research is supported by Sloan Research Fellowship (Award Number FG-2026-79505) from the Alfred P. Sloan Foundation and by an LSST Scialog Early Science grant from the Research Corporation for Science Advancement. This work benefited from interactions supported by the Gordon and Betty Moore Foundation through grant GBMF5076 and through interactions at the Kavli Institute for Theoretical Physics, supported by NSF PHY-2309135.

\bibliographystyle{mnras}
\bibliography{ref.bib}

\begin{thebibliography}{}
\makeatletter
\relax
\def\mn@urlcharsother{\let\do\@makeother \do\$\do\&\do\#\do\^\do\_\do\%\do\~}
\def\mn@doi{\begingroup\mn@urlcharsother \@ifnextchar [ {\mn@doi@}
  {\mn@doi@[]}}
\def\mn@doi@[#1]#2{\def\@tempa{#1}\ifx\@tempa\@empty \href
  {http://dx.doi.org/#2} {doi:#2}\else \href {http://dx.doi.org/#2} {#1}\fi
  \endgroup}
\def\mn@eprint#1#2{\mn@eprint@#1:#2::\@nil}
\def\mn@eprint@arXiv#1{\href {http://arxiv.org/abs/#1} {{\tt arXiv:#1}}}
\def\mn@eprint@dblp#1{\href {http://dblp.uni-trier.de/rec/bibtex/#1.xml}
  {dblp:#1}}
\def\mn@eprint@#1:#2:#3:#4\@nil{\def\@tempa {#1}\def\@tempb {#2}\def\@tempc
  {#3}\ifx \@tempc \@empty \let \@tempc \@tempb \let \@tempb \@tempa \fi \ifx
  \@tempb \@empty \def\@tempb {arXiv}\fi \@ifundefined
  {mn@eprint@\@tempb}{\@tempb:\@tempc}{\expandafter \expandafter \csname
  mn@eprint@\@tempb\endcsname \expandafter{\@tempc}}}

\bibitem[\protect\citeauthoryear{{Aspegren} \& {Kasen}}{{Aspegren} \&
  {Kasen}}{2026}]{2026ApJ..1001..191A}
{Aspegren} O.,  {Kasen} D.,  2026, \mn@doi [\apj] {10.3847/1538-4357/ae5638},
  \href {https://ui.adsabs.harvard.edu/abs/2026ApJ..1001..191A} {1001, 191}

\bibitem[\protect\citeauthoryear{{Blandford} \& {Begelman}}{{Blandford} \&
  {Begelman}}{1999}]{1999MNRAS.303L...1B}
{Blandford} R.~D.,  {Begelman} M.~C.,  1999, \mn@doi [\mnras]
  {10.1046/j.1365-8711.1999.02358.x}, \href
  {https://ui.adsabs.harvard.edu/abs/1999MNRAS.303L...1B} {303, L1}

\bibitem[\protect\citeauthoryear{{Bright} et~al.,}{{Bright}
  et~al.}{2022}]{2022ApJ...926..112B}
{Bright} J.~S.,  et~al., 2022, \mn@doi [\apj] {10.3847/1538-4357/ac4506}, \href
  {https://ui.adsabs.harvard.edu/abs/2022ApJ...926..112B} {926, 112}

\bibitem[\protect\citeauthoryear{{Chen}, {Drout}, {Piro}, {Kilpatrick},
  {Foley}, {Rojas-Bravo}  \& {Magee}}{{Chen}
  et~al.}{2023}]{2023ApJ...955...43C}
{Chen} Y.,  {Drout} M.~R.,  {Piro} A.~L.,  {Kilpatrick} C.~D.,  {Foley} R.~J.,
  {Rojas-Bravo} C.,   {Magee} M.~R.,  2023, \mn@doi [\apj]
  {10.3847/1538-4357/ace964}, \href
  {https://ui.adsabs.harvard.edu/abs/2023ApJ...955...43C} {955, 43}

\bibitem[\protect\citeauthoryear{{Chrimes}, {Jonker}, {Levan}  \&
  {Mummery}}{{Chrimes} et~al.}{2026}]{2026A&A...706A.327C}
{Chrimes} A.~A.,  {Jonker} P.~G.,  {Levan} A.~J.,   {Mummery} A.,  2026,
  \mn@doi [\aap] {10.1051/0004-6361/202557545}, \href
  {https://ui.adsabs.harvard.edu/abs/2026A&A...706A.327C} {706, A327}

\bibitem[\protect\citeauthoryear{{Coppejans} et~al.,}{{Coppejans}
  et~al.}{2020}]{2020ApJ...895L..23C}
{Coppejans} D.~L.,  et~al., 2020, \mn@doi [\apjl] {10.3847/2041-8213/ab8cc7},
  \href {https://ui.adsabs.harvard.edu/abs/2020ApJ...895L..23C} {895, L23}

\bibitem[\protect\citeauthoryear{{Davis}, {Stone}  \& {Pessah}}{{Davis}
  et~al.}{2010}]{2010ApJ...713...52D}
{Davis} S.~W.,  {Stone} J.~M.,   {Pessah} M.~E.,  2010, \mn@doi [\apj]
  {10.1088/0004-637X/713/1/52}, \href
  {https://ui.adsabs.harvard.edu/abs/2010ApJ...713...52D} {713, 52}

\bibitem[\protect\citeauthoryear{{Draine}}{{Draine}}{2011}]{2011piim.book.....D}
{Draine} B.~T.,  2011, {Physics of the Interstellar and Intergalactic Medium}

\bibitem[\protect\citeauthoryear{{Drout} et~al.,}{{Drout}
  et~al.}{2014}]{2014ApJ...794...23D}
{Drout} M.~R.,  et~al., 2014, \mn@doi [\apj] {10.1088/0004-637X/794/1/23},
  \href {https://ui.adsabs.harvard.edu/abs/2014ApJ...794...23D} {794, 23}

\bibitem[\protect\citeauthoryear{{Farmer}, {Renzo}, {de Mink}, {Marchant}  \&
  {Justham}}{{Farmer} et~al.}{2019}]{2019ApJ...887...53F}
{Farmer} R.,  {Renzo} M.,  {de Mink} S.~E.,  {Marchant} P.,   {Justham} S.,
  2019, \mn@doi [\apj] {10.3847/1538-4357/ab518b}, \href
  {https://ui.adsabs.harvard.edu/abs/2019ApJ...887...53F} {887, 53}

\bibitem[\protect\citeauthoryear{{Govreen-Segal}, {Nakar}, {Hotokezaka},
  {Irwin}  \& {Quataert}}{{Govreen-Segal} et~al.}{2026}]{2026arXiv260118887G}
{Govreen-Segal} T.,  {Nakar} E.,  {Hotokezaka} K.,  {Irwin} C.~M.,   {Quataert}
  E.,  2026, \mn@doi [arXiv e-prints] {10.48550/arXiv.2601.18887}, \href
  {https://ui.adsabs.harvard.edu/abs/2026arXiv260118887G} {p. arXiv:2601.18887}

\bibitem[\protect\citeauthoryear{{Guo}, {Stone}, {Quataert}  \& {Kim}}{{Guo}
  et~al.}{2024}]{2024ApJ...973..141G}
{Guo} M.,  {Stone} J.~M.,  {Quataert} E.,   {Kim} C.-G.,  2024, \mn@doi [\apj]
  {10.3847/1538-4357/ad5fe7}, \href
  {https://ui.adsabs.harvard.edu/abs/2024ApJ...973..141G} {973, 141}

\bibitem[\protect\citeauthoryear{{Hawley}, {Guan}  \& {Krolik}}{{Hawley}
  et~al.}{2011}]{2011ApJ...738...84H}
{Hawley} J.~F.,  {Guan} X.,   {Krolik} J.~H.,  2011, \mn@doi [\apj]
  {10.1088/0004-637X/738/1/84}, \href
  {https://ui.adsabs.harvard.edu/abs/2011ApJ...738...84H} {738, 84}

\bibitem[\protect\citeauthoryear{{Ho} et~al.,}{{Ho}
  et~al.}{2019}]{2019ApJ...871...73H}
{Ho} A. Y.~Q.,  et~al., 2019, \mn@doi [\apj] {10.3847/1538-4357/aaf473}, \href
  {https://ui.adsabs.harvard.edu/abs/2019ApJ...871...73H} {871, 73}

\bibitem[\protect\citeauthoryear{{Ho} et~al.,}{{Ho}
  et~al.}{2020}]{2020ApJ...895...49H}
{Ho} A. Y.~Q.,  et~al., 2020, \mn@doi [\apj] {10.3847/1538-4357/ab8bcf}, \href
  {https://ui.adsabs.harvard.edu/abs/2020ApJ...895...49H} {895, 49}

\bibitem[\protect\citeauthoryear{{Ho} et~al.,}{{Ho}
  et~al.}{2023}]{2023Natur.623..927H}
{Ho} A. Y.~Q.,  et~al., 2023, \mn@doi [\nat] {10.1038/s41586-023-06673-6},
  \href {https://ui.adsabs.harvard.edu/abs/2023Natur.623..927H} {623, 927}

\bibitem[\protect\citeauthoryear{{Inkenhaag}, {Jonker}, {Levan}, {Chrimes},
  {Mummery}, {Perley}  \& {Tanvir}}{{Inkenhaag}
  et~al.}{2023}]{2023MNRAS.525.4042I}
{Inkenhaag} A.,  {Jonker} P.~G.,  {Levan} A.~J.,  {Chrimes} A.~A.,  {Mummery}
  A.,  {Perley} D.~A.,   {Tanvir} N.~R.,  2023, \mn@doi [\mnras]
  {10.1093/mnras/stad2531}, \href
  {https://ui.adsabs.harvard.edu/abs/2023MNRAS.525.4042I} {525, 4042}

\bibitem[\protect\citeauthoryear{{Inkenhaag}, {Levan}, {Mummery}  \&
  {Jonker}}{{Inkenhaag} et~al.}{2025}]{2025MNRAS.544L.108I}
{Inkenhaag} A.,  {Levan} A.~J.,  {Mummery} A.,   {Jonker} P.~G.,  2025, \mn@doi
  [\mnras] {10.1093/mnrasl/slaf107}, \href
  {https://ui.adsabs.harvard.edu/abs/2025MNRAS.544L.108I} {544, L108}

\bibitem[\protect\citeauthoryear{{Klencki} \& {Metzger}}{{Klencki} \&
  {Metzger}}{2025}]{2025arXiv251009745K}
{Klencki} J.,  {Metzger} B.~D.,  2025, \mn@doi [arXiv e-prints]
  {10.48550/arXiv.2510.09745}, \href
  {https://ui.adsabs.harvard.edu/abs/2025arXiv251009745K} {p. arXiv:2510.09745}

\bibitem[\protect\citeauthoryear{{Kremer}, {Lu}, {Rodriguez}, {Lachat}  \&
  {Rasio}}{{Kremer} et~al.}{2019}]{2019ApJ...881...75K}
{Kremer} K.,  {Lu} W.,  {Rodriguez} C.~L.,  {Lachat} M.,   {Rasio} F.~A.,
  2019, \mn@doi [\apj] {10.3847/1538-4357/ab2e0c}, \href
  {https://ui.adsabs.harvard.edu/abs/2019ApJ...881...75K} {881, 75}

\bibitem[\protect\citeauthoryear{{Kremer}, {Lombardi}, {Lu}, {Piro}  \&
  {Rasio}}{{Kremer} et~al.}{2022}]{2022ApJ...933..203K}
{Kremer} K.,  {Lombardi} J.~C.,  {Lu} W.,  {Piro} A.~L.,   {Rasio} F.~A.,
  2022, \mn@doi [\apj] {10.3847/1538-4357/ac714f}, \href
  {https://ui.adsabs.harvard.edu/abs/2022ApJ...933..203K} {933, 203}

\bibitem[\protect\citeauthoryear{{Kuin} et~al.,}{{Kuin}
  et~al.}{2019}]{2019MNRAS.487.2505K}
{Kuin} N. P.~M.,  et~al., 2019, \mn@doi [\mnras] {10.1093/mnras/stz053}, \href
  {https://ui.adsabs.harvard.edu/abs/2019MNRAS.487.2505K} {487, 2505}

\bibitem[\protect\citeauthoryear{{LeBaron} et~al.,}{{LeBaron}
  et~al.}{2026}]{2026ApJ...997L..10L}
{LeBaron} N.,  et~al., 2026, \mn@doi [\apjl] {10.3847/2041-8213/ae2910}, \href
  {https://ui.adsabs.harvard.edu/abs/2026ApJ...997L..10L} {997, L10}

\bibitem[\protect\citeauthoryear{{Liska}, {Tchekhovskoy}  \&
  {Quataert}}{{Liska} et~al.}{2020}]{2020MNRAS.494.3656L}
{Liska} M.,  {Tchekhovskoy} A.,   {Quataert} E.,  2020, \mn@doi [\mnras]
  {10.1093/mnras/staa955}, \href
  {https://ui.adsabs.harvard.edu/abs/2020MNRAS.494.3656L} {494, 3656}

\bibitem[\protect\citeauthoryear{{Lu}}{{Lu}}{2022}]{2022hxga.book....3L}
{Lu} W.,  2022, in {Bambi} C.,  {Sangangelo} A.,  eds, , Handbook of X-ray and
  Gamma-ray Astrophysics.
p.~3, \mn@doi{10.1007/978-981-16-4544-0_127-1}

\bibitem[\protect\citeauthoryear{{Lu}, {Fuller}, {Quataert}  \&
  {Bonnerot}}{{Lu} et~al.}{2023}]{2023MNRAS.519.1409L}
{Lu} W.,  {Fuller} J.,  {Quataert} E.,   {Bonnerot} C.,  2023, \mn@doi [\mnras]
  {10.1093/mnras/stac3621}, \href
  {https://ui.adsabs.harvard.edu/abs/2023MNRAS.519.1409L} {519, 1409}

\bibitem[\protect\citeauthoryear{{Lyutikov} \& {Toonen}}{{Lyutikov} \&
  {Toonen}}{2019}]{2019MNRAS.487.5618L}
{Lyutikov} M.,  {Toonen} S.,  2019, \mn@doi [\mnras] {10.1093/mnras/stz1640},
  \href {https://ui.adsabs.harvard.edu/abs/2019MNRAS.487.5618L} {487, 5618}

\bibitem[\protect\citeauthoryear{{Margutti} et~al.,}{{Margutti}
  et~al.}{2019}]{2019ApJ...872...18M}
{Margutti} R.,  et~al., 2019, \mn@doi [\apj] {10.3847/1538-4357/aafa01}, \href
  {https://ui.adsabs.harvard.edu/abs/2019ApJ...872...18M} {872, 18}

\bibitem[\protect\citeauthoryear{{Metzger}}{{Metzger}}{2022}]{2022ApJ...932...84M}
{Metzger} B.~D.,  2022, \mn@doi [\apj] {10.3847/1538-4357/ac6d59}, \href
  {https://ui.adsabs.harvard.edu/abs/2022ApJ...932...84M} {932, 84}

\bibitem[\protect\citeauthoryear{{Migliori} et~al.,}{{Migliori}
  et~al.}{2024}]{2024ApJ...963L..24M}
{Migliori} G.,  et~al., 2024, \mn@doi [\apjl] {10.3847/2041-8213/ad2764}, \href
  {https://ui.adsabs.harvard.edu/abs/2024ApJ...963L..24M} {963, L24}

\bibitem[\protect\citeauthoryear{{Morokuma-Matsui} et~al.,}{{Morokuma-Matsui}
  et~al.}{2019}]{2019ApJ...879L..13M}
{Morokuma-Matsui} K.,  et~al., 2019, \mn@doi [\apjl]
  {10.3847/2041-8213/ab2915}, \href
  {https://ui.adsabs.harvard.edu/abs/2019ApJ...879L..13M} {879, L13}

\bibitem[\protect\citeauthoryear{{Nayana} et~al.,}{{Nayana}
  et~al.}{2025}]{2025ApJ...993L...6N}
{Nayana} A.~J.,  et~al., 2025, \mn@doi [\apjl] {10.3847/2041-8213/ae0b4d},
  \href {https://ui.adsabs.harvard.edu/abs/2025ApJ...993L...6N} {993, L6}

\bibitem[\protect\citeauthoryear{{Perets}, {Li}, {Lombardi}  \&
  {Milcarek}}{{Perets} et~al.}{2016}]{2016ApJ...823..113P}
{Perets} H.~B.,  {Li} Z.,  {Lombardi} Jr. J.~C.,   {Milcarek} Jr. S.~R.,  2016,
  \mn@doi [\apj] {10.3847/0004-637X/823/2/113}, \href
  {https://ui.adsabs.harvard.edu/abs/2016ApJ...823..113P} {823, 113}

\bibitem[\protect\citeauthoryear{{Perley} et~al.,}{{Perley}
  et~al.}{2019}]{2019MNRAS.484.1031P}
{Perley} D.~A.,  et~al., 2019, \mn@doi [\mnras] {10.1093/mnras/sty3420}, \href
  {https://ui.adsabs.harvard.edu/abs/2019MNRAS.484.1031P} {484, 1031}

\bibitem[\protect\citeauthoryear{{Perley} et~al.,}{{Perley}
  et~al.}{2021}]{2021MNRAS.508.5138P}
{Perley} D.~A.,  et~al., 2021, \mn@doi [\mnras] {10.1093/mnras/stab2785}, \href
  {https://ui.adsabs.harvard.edu/abs/2021MNRAS.508.5138P} {508, 5138}

\bibitem[\protect\citeauthoryear{{Prentice} et~al.,}{{Prentice}
  et~al.}{2018}]{2018ApJ...865L...3P}
{Prentice} S.~J.,  et~al., 2018, \mn@doi [\apjl] {10.3847/2041-8213/aadd90},
  \href {https://ui.adsabs.harvard.edu/abs/2018ApJ...865L...3P} {865, L3}

\bibitem[\protect\citeauthoryear{{Quataert}, {Fern{\'a}ndez}, {Kasen}, {Klion}
  \& {Paxton}}{{Quataert} et~al.}{2016}]{2016MNRAS.458.1214Q}
{Quataert} E.,  {Fern{\'a}ndez} R.,  {Kasen} D.,  {Klion} H.,   {Paxton} B.,
  2016, \mn@doi [\mnras] {10.1093/mnras/stw365}, \href
  {https://ui.adsabs.harvard.edu/abs/2016MNRAS.458.1214Q} {458, 1214}

\bibitem[\protect\citeauthoryear{{Sevilla} et~al.,}{{Sevilla}
  et~al.}{2026}]{2026arXiv260118926S}
{Sevilla} C.,  et~al., 2026, \mn@doi [arXiv e-prints]
  {10.48550/arXiv.2601.18926}, \href
  {https://ui.adsabs.harvard.edu/abs/2026arXiv260118926S} {p. arXiv:2601.18926}

\bibitem[\protect\citeauthoryear{{Shakura} \& {Sunyaev}}{{Shakura} \&
  {Sunyaev}}{1973}]{1973A&A....24..337S}
{Shakura} N.~I.,  {Sunyaev} R.~A.,  1973, \aap, \href
  {https://ui.adsabs.harvard.edu/abs/1973A&A....24..337S} {24, 337}

\bibitem[\protect\citeauthoryear{{Shen}, {Nakar}  \& {Piran}}{{Shen}
  et~al.}{2016}]{2016MNRAS.459..171S}
{Shen} R.-F.,  {Nakar} E.,   {Piran} T.,  2016, \mn@doi [\mnras]
  {10.1093/mnras/stw645}, \href
  {https://ui.adsabs.harvard.edu/abs/2016MNRAS.459..171S} {459, 171}

\bibitem[\protect\citeauthoryear{{Soker}, {Grichener}  \& {Gilkis}}{{Soker}
  et~al.}{2019}]{2019MNRAS.484.4972S}
{Soker} N.,  {Grichener} A.,   {Gilkis} A.,  2019, \mn@doi [\mnras]
  {10.1093/mnras/stz364}, \href
  {https://ui.adsabs.harvard.edu/abs/2019MNRAS.484.4972S} {484, 4972}

\bibitem[\protect\citeauthoryear{{Somalwar} et~al.,}{{Somalwar}
  et~al.}{2025}]{2025ApJ...995..228S}
{Somalwar} J.~J.,  et~al., 2025, \mn@doi [\apj] {10.3847/1538-4357/ae1501},
  \href {https://ui.adsabs.harvard.edu/abs/2025ApJ...995..228S} {995, 228}

\bibitem[\protect\citeauthoryear{{Speagle}}{{Speagle}}{2020}]{2020MNRAS.493.3132S}
{Speagle} J.~S.,  2020, \mn@doi [\mnras] {10.1093/mnras/staa278}, \href
  {https://ui.adsabs.harvard.edu/abs/2020MNRAS.493.3132S} {493, 3132}

\bibitem[\protect\citeauthoryear{{Stone}, {Pringle}  \& {Begelman}}{{Stone}
  et~al.}{1999}]{1999MNRAS.310.1002S}
{Stone} J.~M.,  {Pringle} J.~E.,   {Begelman} M.~C.,  1999, \mn@doi [\mnras]
  {10.1046/j.1365-8711.1999.03024.x}, \href
  {https://ui.adsabs.harvard.edu/abs/1999MNRAS.310.1002S} {310, 1002}

\bibitem[\protect\citeauthoryear{{Sun}, {Maund}, {Crowther}  \& {Liu}}{{Sun}
  et~al.}{2022}]{2022MNRAS.512L..66S}
{Sun} N.-C.,  {Maund} J.~R.,  {Crowther} P.~A.,   {Liu} L.-D.,  2022, \mn@doi
  [\mnras] {10.1093/mnrasl/slac023}, \href
  {https://ui.adsabs.harvard.edu/abs/2022MNRAS.512L..66S} {512, L66}

\bibitem[\protect\citeauthoryear{{Sun}, {Maund}, {Shao}  \& {Janiak}}{{Sun}
  et~al.}{2023}]{2023MNRAS.519.3785S}
{Sun} N.-C.,  {Maund} J.~R.,  {Shao} Y.,   {Janiak} I.~A.,  2023, \mn@doi
  [\mnras] {10.1093/mnras/stac3773}, \href
  {https://ui.adsabs.harvard.edu/abs/2023MNRAS.519.3785S} {519, 3785}

\bibitem[\protect\citeauthoryear{{Tsuna} \& {Lu}}{{Tsuna} \&
  {Lu}}{2025}]{2025ApJ...986...84T}
{Tsuna} D.,  {Lu} W.,  2025, \mn@doi [\apj] {10.3847/1538-4357/add158}, \href
  {https://ui.adsabs.harvard.edu/abs/2025ApJ...986...84T} {986, 84}

\bibitem[\protect\citeauthoryear{{Vynatheya}, {Ryu}, {Pakmor}, {de Mink}  \&
  {Perets}}{{Vynatheya} et~al.}{2024}]{2024A&A...685A..45V}
{Vynatheya} P.,  {Ryu} T.,  {Pakmor} R.,  {de Mink} S.~E.,   {Perets} H.~B.,
  2024, \mn@doi [\aap] {10.1051/0004-6361/202348357}, \href
  {https://ui.adsabs.harvard.edu/abs/2024A&A...685A..45V} {685, A45}

\bibitem[\protect\citeauthoryear{{Winter-Granic} \& {Quataert}}{{Winter-Granic}
  \& {Quataert}}{2025}]{2025arXiv251209017W}
{Winter-Granic} M.,  {Quataert} E.,  2025, \mn@doi [arXiv e-prints]
  {10.48550/arXiv.2512.09017}, \href
  {https://ui.adsabs.harvard.edu/abs/2025arXiv251209017W} {p. arXiv:2512.09017}

\bibitem[\protect\citeauthoryear{{Woosley}}{{Woosley}}{2017}]{2017ApJ...836..244W}
{Woosley} S.~E.,  2017, \mn@doi [\apj] {10.3847/1538-4357/836/2/244}, \href
  {https://ui.adsabs.harvard.edu/abs/2017ApJ...836..244W} {836, 244}

\bibitem[\protect\citeauthoryear{{Wright} \& {Barlow}}{{Wright} \&
  {Barlow}}{1975}]{1975MNRAS.170...41W}
{Wright} A.~E.,  {Barlow} M.~J.,  1975, \mn@doi [\mnras]
  {10.1093/mnras/170.1.41}, \href
  {https://ui.adsabs.harvard.edu/abs/1975MNRAS.170...41W} {170, 41}

\bibitem[\protect\citeauthoryear{{Yao} et~al.,}{{Yao}
  et~al.}{2022}]{2022ApJ...934..104Y}
{Yao} Y.,  et~al., 2022, \mn@doi [\apj] {10.3847/1538-4357/ac7a41}, \href
  {https://ui.adsabs.harvard.edu/abs/2022ApJ...934..104Y} {934, 104}

\bibitem[\protect\citeauthoryear{{Yuan} \& {Narayan}}{{Yuan} \&
  {Narayan}}{2014}]{2014ARA&A..52..529Y}
{Yuan} F.,  {Narayan} R.,  2014, \mn@doi [\araa]
  {10.1146/annurev-astro-082812-141003}, \href
  {https://ui.adsabs.harvard.edu/abs/2014ARA&A..52..529Y} {52, 529}

\makeatother
\end{thebibliography}

\appendix

\begin{figure*}
\centering
\includegraphics[width=0.47\textwidth]{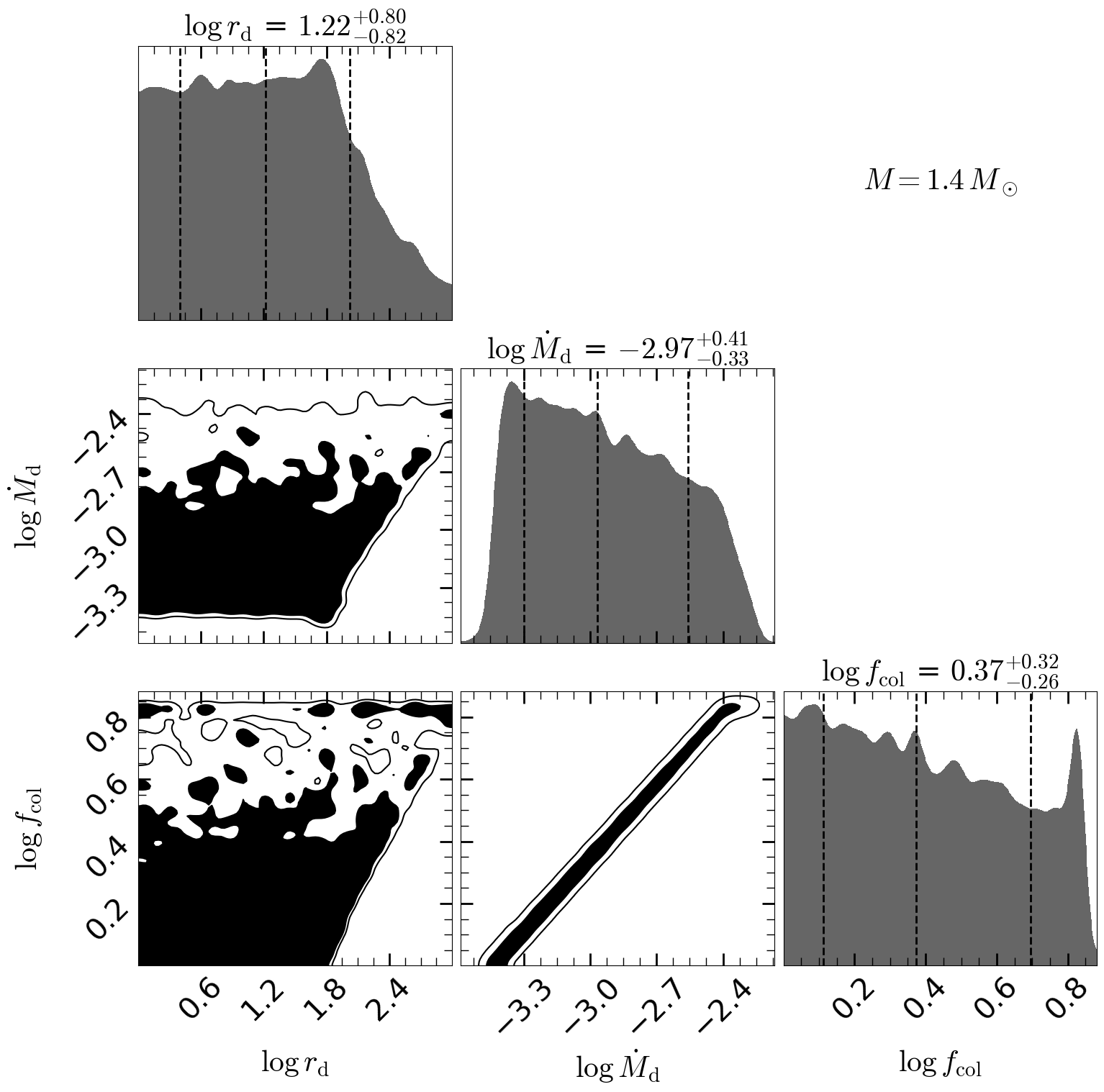}
\includegraphics[width=0.47\textwidth]{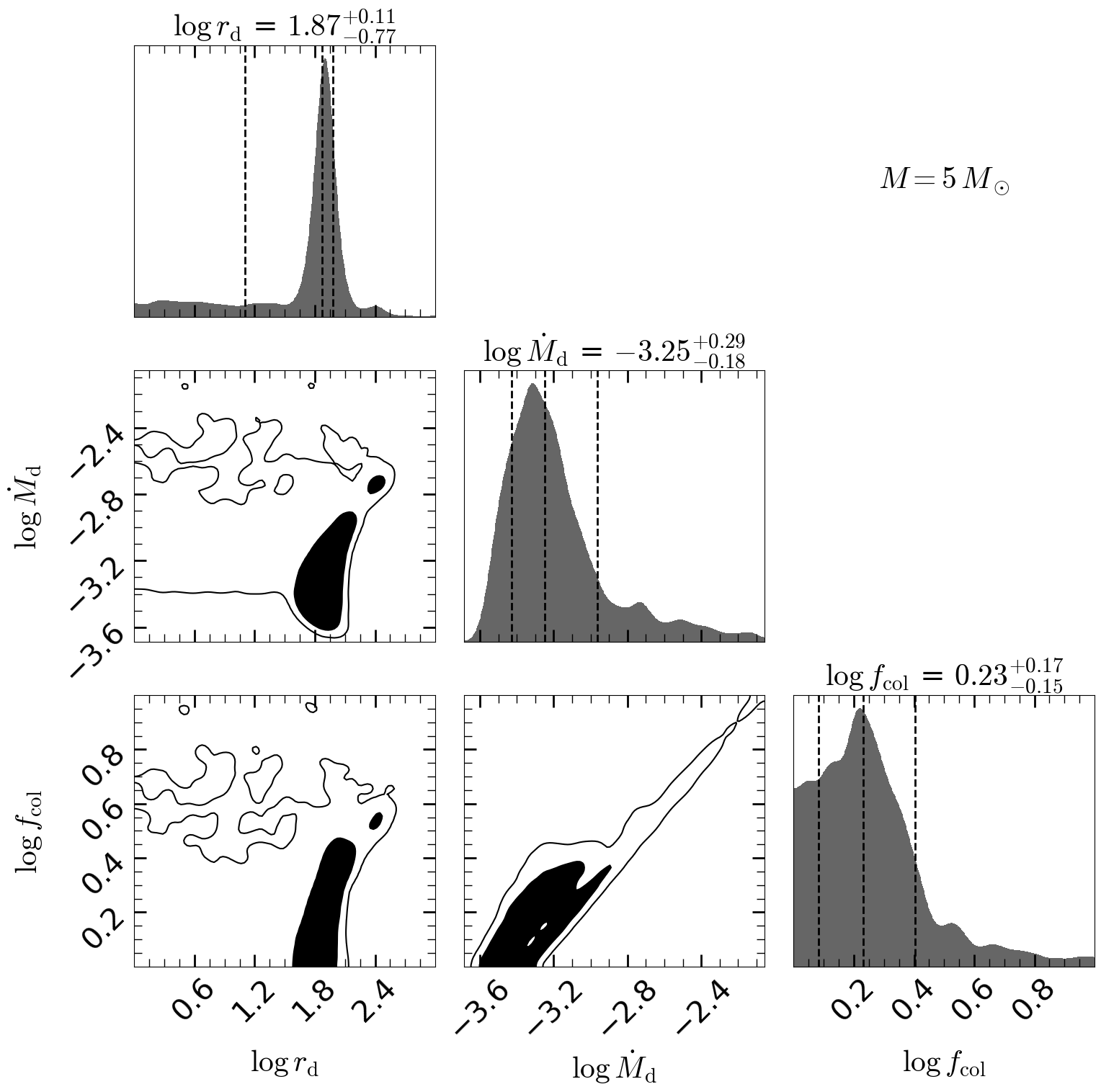}
\caption{Corner plots for the posterior distributions of our disk+wind model as fitted to the AT2018cow HST data at $t=1453\rm\, d$. Disk outer radius $r_{\rm d}$ is in units of $R_\odot$ and accretion rate $\dot{M}_{\rm d}$ is in units of $\msunyr$. Two different compact object masses are considered: $M=1.4M_\odot$ (left panel) and $5M_\odot$ (right panel).
}
\label{fig:corner_M1p4_M5}
\end{figure*}

\section{A. Full posteriors}\label{app:full_posteriors}

\begin{figure*}
\centering
\includegraphics[width=0.47\textwidth]{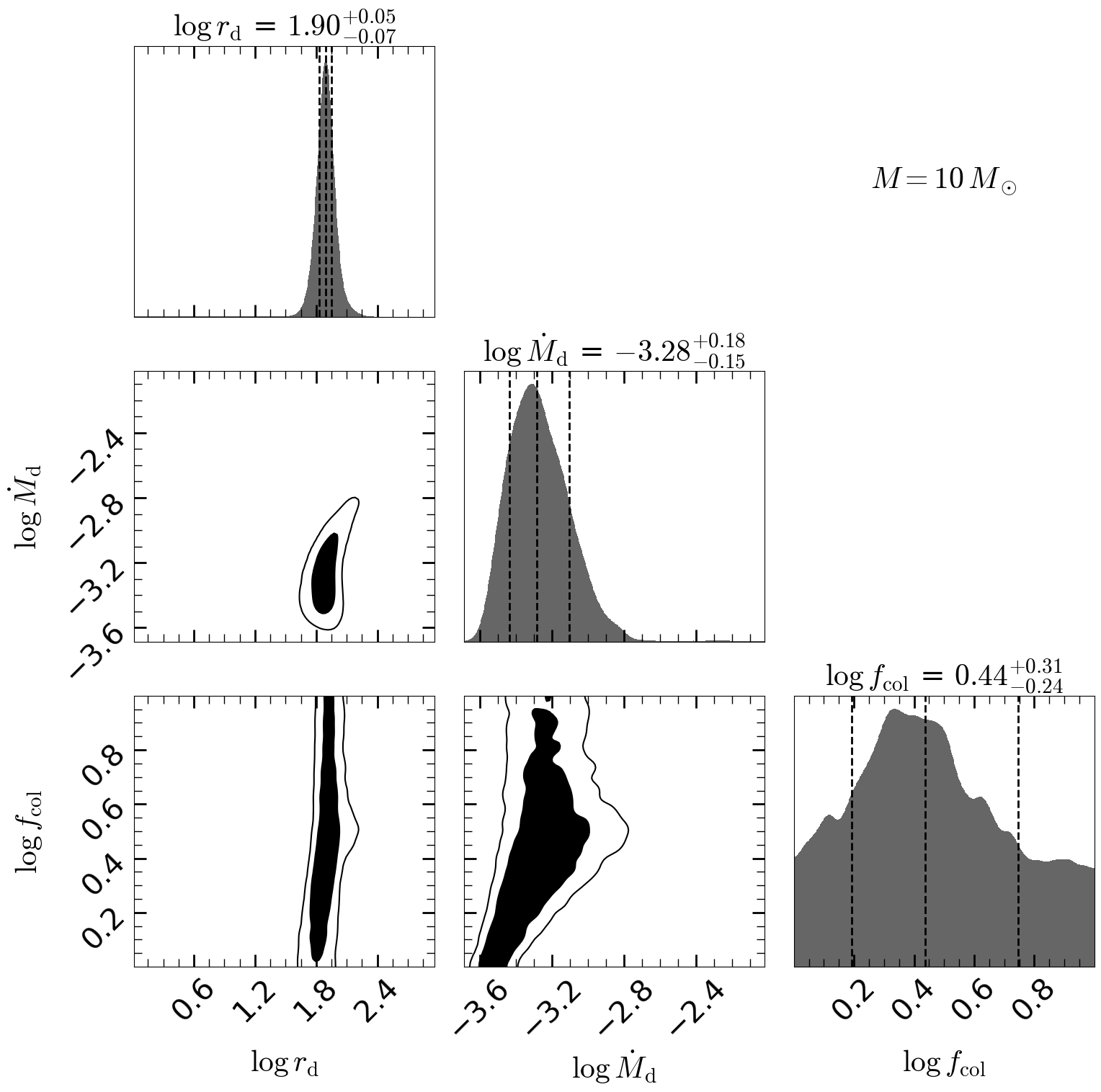}
\includegraphics[width=0.47\textwidth]{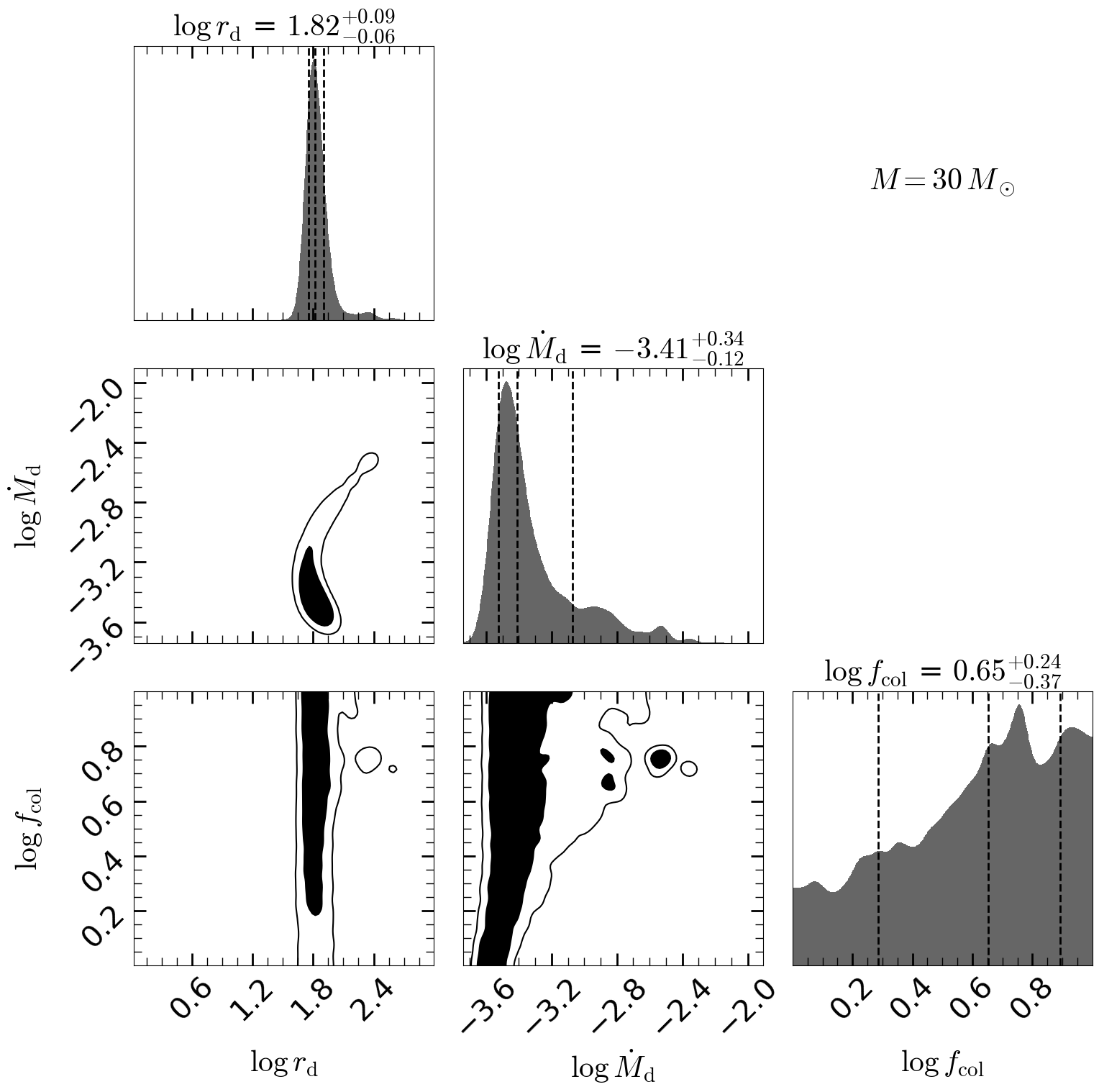}
\caption{Corner plots for $M=10M_\odot$ and $30M_\odot$.
}
\label{fig:corner_M10_M30}
\end{figure*}

\begin{figure*}
\centering
\includegraphics[width=0.47\textwidth]{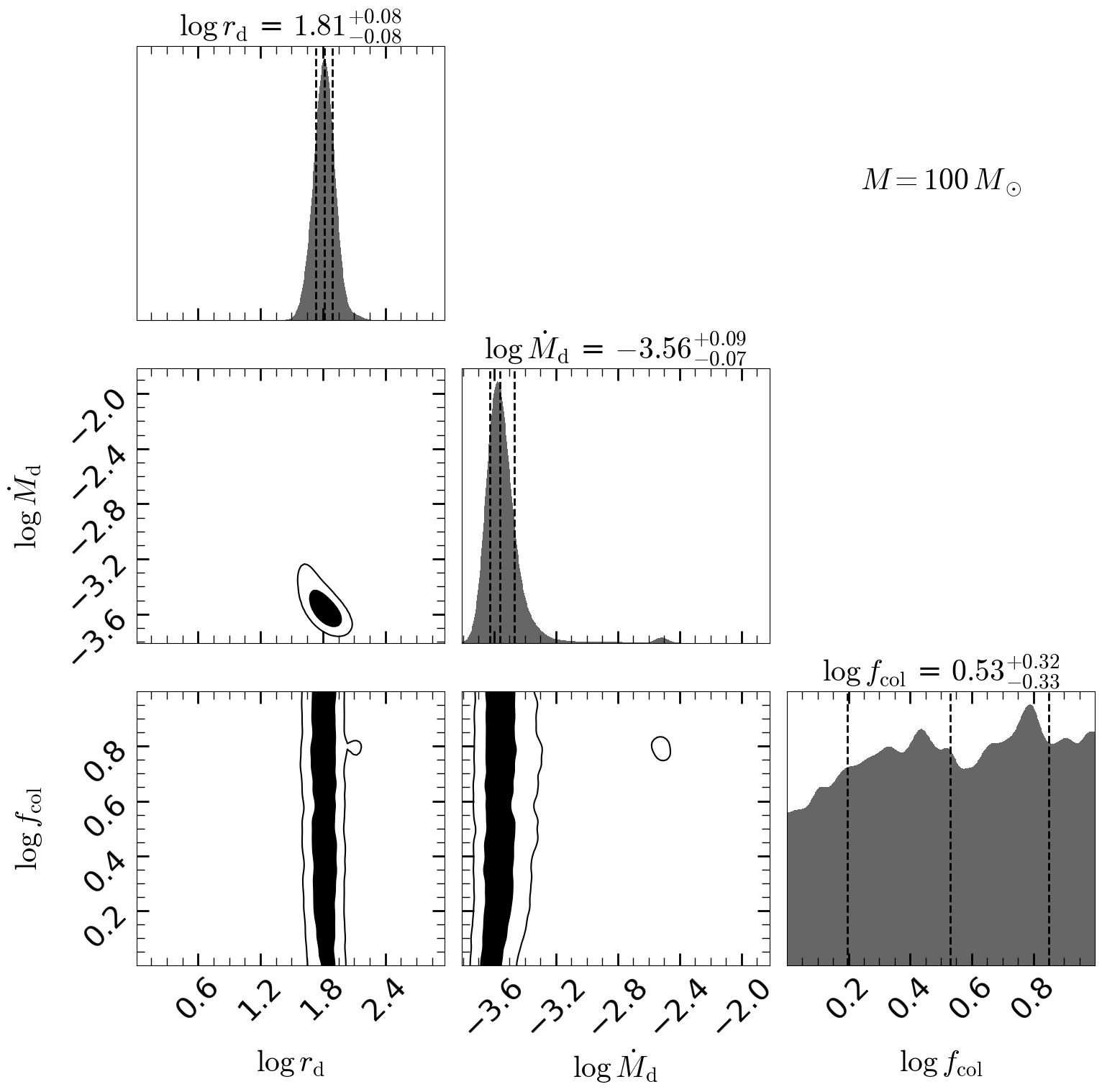}
\includegraphics[width=0.47\textwidth]{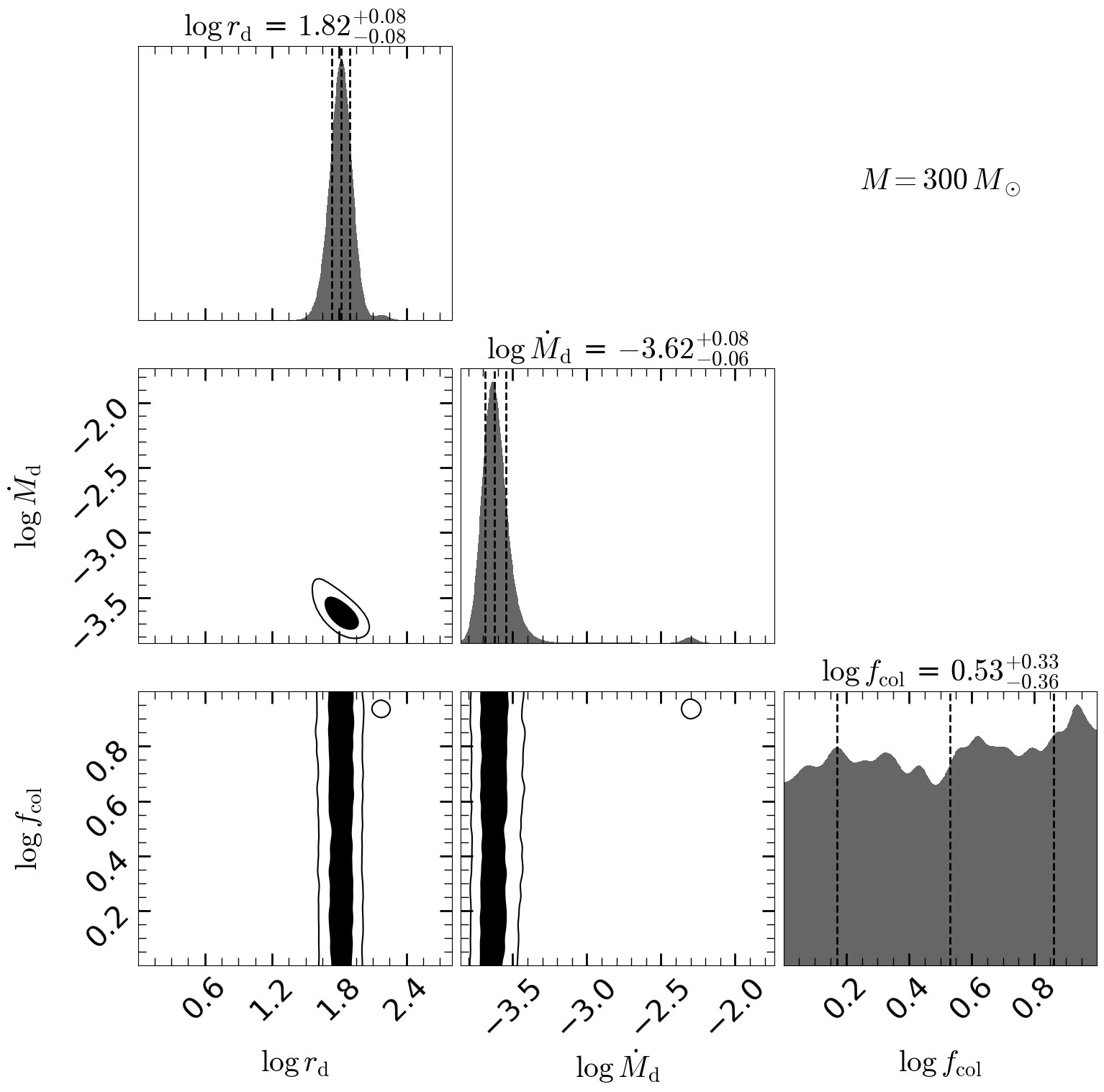}
\caption{Corner plots for $M=100M_\odot$ and $300M_\odot$.
}
\label{fig:corner_M100_M300}
\end{figure*}

\begin{figure}
\centering
\includegraphics[width=0.47\textwidth]{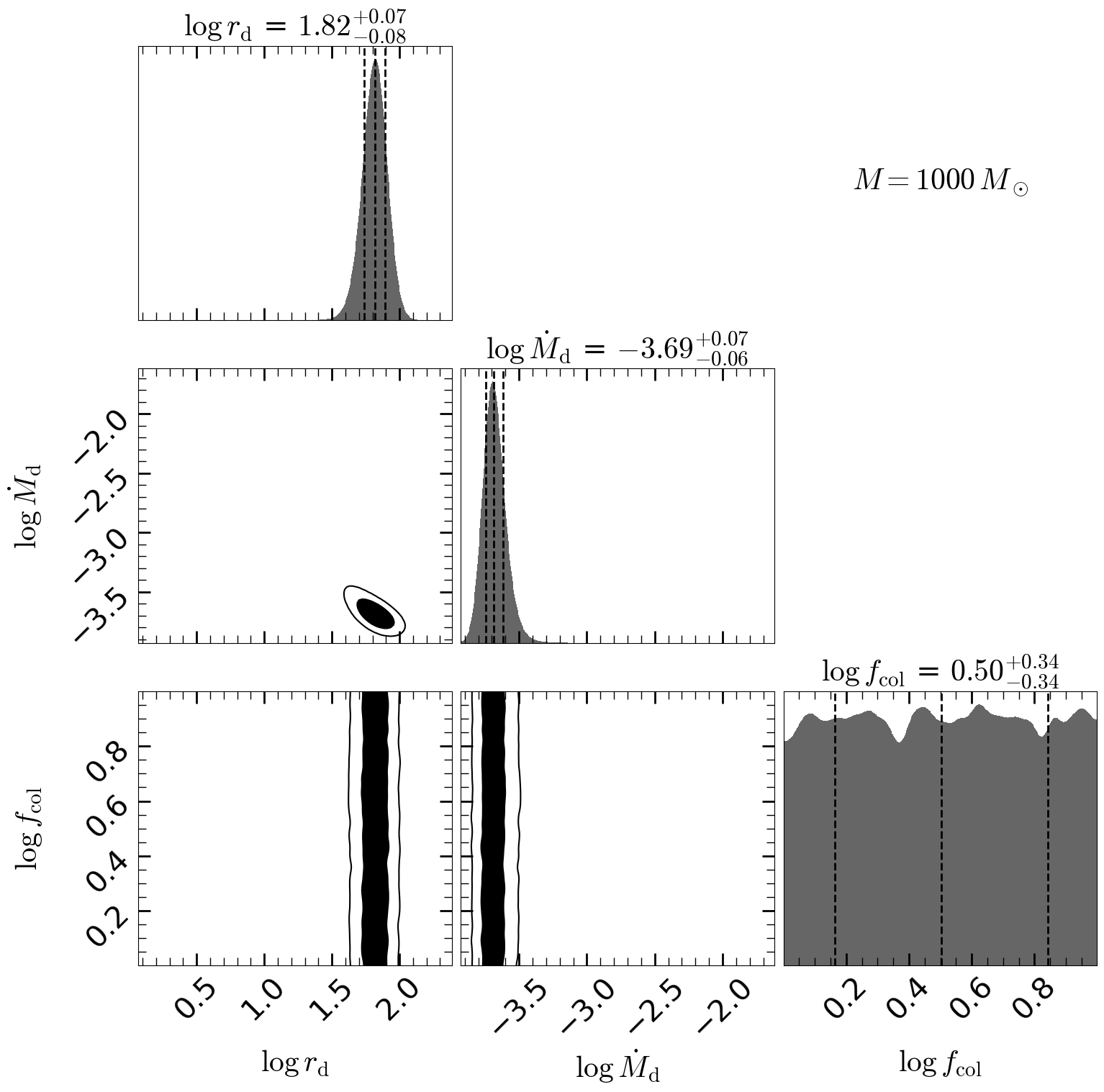}
\caption{Corner plot for $M=1000M_\odot$.
}
\label{fig:corner_M1000}
\end{figure}

The full posterior distributions for each of the compact object masses are shown in Figs. \ref{fig:corner_M1p4_M5}, \ref{fig:corner_M10_M30}, \ref{fig:corner_M100_M300}, and \ref{fig:corner_M1000}.

\end{document}